\documentclass[12pt]{article}
\usepackage{a4wide}
\usepackage{graphics}
\usepackage{amsfonts}
\usepackage{amstext}
\usepackage{amsmath}
\usepackage{amssymb}
\usepackage{amsthm}
\usepackage{epic,eepic}
\usepackage{psfrag}
\usepackage{youngtab}
\usepackage{cite}

\renewcommand{\theequation}{\arabic{section}.\arabic{equation}}

\def\Ai{{\rm Ai}}
\def\b{\beta}
\def\G{\Gamma}
\def\K{\mathcal{K}}
\def\vp{\varphi}
\def\t{\tau}

\begin{document}
\title{Fluctuations of the one-dimensional asymmetric 
exclusion process using random matrix techniques}
\author{
\vspace{5mm}
T. Sasamoto
{\footnote {\tt email: sasamoto@math.s.chiba-u.ac.jp}}
\\
{\it Department of Mathematics and Informatics, Chiba University}\\
\vspace{5mm}
{\it 1-33 Yayoi-cho, Inage, Chiba, 263-8522, Japan}\\
}
\maketitle

\begin{abstract}

The studies of fluctuations of the one-dimensional
Kardar-Parisi-Zhang universality class using the 
techniques from random matrix theory are reviewed 
from the point of view of the asymmetric simple 
exclusion process. 
We explain the basics of random matrix techniques,
the connections to the polynuclear growth models 
and a method using the Green's function. 
\end{abstract}

\setcounter{equation}{0}
\section{Introduction}
The one-dimensional asymmetric simple exclusion process
(ASEP) \cite{Liggett1985,Liggett1999} 
is a stochastic process of many particles which 
perform asymmetric random walks with hardcore exclusion 
interaction. 
The ASEP is a simple model of statistical mechanical systems 
driven away from equilibrium \cite{Spohn1991,Schuetz2000}. 
In spite of its simplicity, it shows a rich variety of 
phenomena and has attracted much attention. 
The stationary properties at a single time are 
already of great physical interest.
A lot of insights have been obtained about the 
boundary induced phase transitions, the long range 
correlations and so on using the powerful 
matrix product method \cite{DEHP1993}.

As for time dependent properties of the ASEP, the average 
behaviors are rather well described by the mean-field 
approximation. We can also understand some part of 
fluctuations from the fact that the ASEP belongs to the 
Kardar-Parisi-Zhang(KPZ) universality class of surface growth 
models \cite{KPZ1986,BS1995,KS1992}. 
In particular the critical exponents are understood by 
the dynamical renormalization method and can be confirmed by some 
exact solutions \cite{GS1992,Ki1995}.

In the past decade we have seen a substantial progress of
understanding of this scaling behaviors of the 1D KPZ systems. 
In the language of the ASEP, a different approach based on 
a connection to the random matrix theory has been 
developed for the infinite system. 
It has allowed us to compute several time dependent quantities 
quite explicitly. Several important results and insights which 
have been obtained so far are 
(1)The fluctuations of the integrated current for several initial 
conditions have been computed and are shown to be equivalent to those of 
the largest eigenvalue of Gaussian ensembles. 
(2) The stationary two point correlation function has been computed 
explicitly. 
(3) The effects of boundary and initial conditions have been
made quite explicit for certain special cases. 

A problem is that these results are rather involved and often 
described in the language of random matrix theory. Moreover 
some results are stated from the point of view of other 
related models and the connection to the ASEP is not obvious.
Hence for those who are not very familiar with random 
matrix theory and these models, the developments might 
look not easily accessible. 
The purpose of this article is to explain the basics of random 
matrix techniques and the relationship to a few of the related 
models. Some part of this paper will overlap with 
the paper by Pr\"ahofer and Spohn \cite{PS2000b} which 
contains a lot of information on the same topic. 
The present article aims at giving some background 
knowledge for \cite{PS2002b} and at the same time a collection 
of new results obtained after \cite{PS2002b}.
See also recent review papers on related topics
\cite{Johansson2006,Spohn2006,Deift2006p,FerrariPraehofer2006}.

The article is organized as follows. In the next section,
we define the model. We mainly study the discrete time
TASEP with parallel update scheme. The continuous time version 
is also considered. 
In section 3, we map the problem of current to a combinatorial 
one and get an expression in the form of multiple integrals. 
In section 4, the basics of the random matrix techniques are
explained. 
The multi-point fluctuations are studied in section 5 by 
introducing the discrete polynuclear growth model and its 
multi-layer version.  
Some generalizations and variants are explained in 
section 6.
In section 7, we explain another approach based on the 
Green's function. 
The final section is devoted to the concluding remarks.

\setcounter{equation}{0}
\section{Models}
In this article we mainly consider a discrete time 
TASEP (dTASEP) on the infinite lattice, $\mathbb{Z}$. There 
are several versions of the update schemes. Here
we employ the parallel update for which the model is 
defined as follows. Each site of the lattice can be occupied 
by a particle or is empty (Fig. 1). Suppose that at each time 
step each particle tosses its own coin which shows heads with 
probability $1-q$. If it turns out a head the particle tries to 
hop to the right neighboring site.
If the target site is occupied, the hopping does not occur 
due to the exclusion interaction among particles.
If it turns out a tail, on the other hand, the particle remains
at the same site. This is the TASEP with parallel update. 
The parameter $q$ here should not be confused with 
the asymmetry parameter in the partially ASEP which 
is often referred to as $q$ in the study of ASEP. 

The above definition describes how the configuration of 
the lattice changes as time goes on. It is also possible 
to define the same process by focusing on when each particle 
makes jumps to the right. Notice that the distribution of the 
waiting time of each hopping after it becomes possible 
is geometrically distributed with parameter $q$.
Suppose that each particle has its own clock instead of the coin. 
It rings after geometrically distributed time at which the particle 
tries to hop to the right neighboring site. If the target 
site is occupied, the hopping does not occur. 
This description of the process defines the 
same TASEP with parallel update as above. 

The stationary measure with constant density $\rho$ is 
rather simple. It is determined by the condition that 
the probability $P(\eta_j, \eta_{j+1})$ that arbitrary 
consecutive two sites $j,j+1$ are occupied ($\eta_j=1$)
or empty ($\eta_j=0$) is given by \cite{JPS1995p} 
\begin{align}
 P(00) &= 1-\rho-J/(1-q), \\
 P(01) &= P(10) = J/(1-q), \\
 P(11) &= \rho-J/(1-q),
\end{align}
where $J$ is the average current, 
\begin{equation}
 J = \frac12 (1-\sqrt{1-4(1-q)\rho(1-\rho)}).
\end{equation}
We remark that in the stationary state the waiting time 
of each particle before making a hop is geometrically 
distributed with parameter $1-J/\rho$.

In the study of nonequilibrium systems, one of the most 
fundamental quantity is the current. 
Let $N(t)$ be the number of particles which crossed the 
bond between sites 0 and 1 from time 0 through $t$.
The average behavior is easily seen to be $N(t) \sim Jt$, 
or more precisely, 
\begin{equation}
 \lim_{t\to\infty} \frac{N(t)}{t} = J.
\end{equation}
Hence the next step is to understand the fluctuation 
of this quantity. From the KPZ theory the exponent
of the fluctuation is expected to scale as
$N(t)-Jt \sim O(t^{1/3})$. The random matrix 
techniques allow us to compute even the scaling function. 
Note that the stationary measure is not 
enough to get such information. 

There is an important limiting case; $q\to 1$.
Since the hopping probability $1-q$ goes to zero, one has to 
rescale the time at the same time to take a meaningful limit. 
Let us consider the limit $q\to 1, t\to \infty$ with 
$\tilde{t}=(1-q)t$ fixed. 
In this limit the process reduces to the continuous time TASEP 
(cTASEP) which is the most well studied in the study of ASEP 
\cite{Liggett1985,Liggett1999}. In cTASEP during infinitesimal 
time duration $dt$, each particle tries to hop to the right 
neighboring site with probability $dt$ under the same exclusion 
principle as for the discrete case. It is again possible to 
define the model using the waiting time for hoppings. 
The only difference as compared to the dTASEP is that each 
clock now rings after an exponential time with parameter one. 
A remark is in order. In the following, when the cTASEP is 
discussed, the time $t$ should be understood as the rescaled 
time $\tilde{t}$.

The stationary measure of the cTASEP with density $\rho$ is 
Bernoulli with parameter $\rho$, i.e., each site is independent 
and is occupied with probability $\rho$  or is empty with probability
$1-\rho$. The average current is known to be $J=\rho(1-\rho)$.

\setcounter{equation}{0}
\section{Combinatorial Approach}
In this section we explain a result for the fluctuation 
of $N(t)$ following \cite{Johansson2000}. 
The initial condition is taken to be the step one, 
in which all sites $x \leq 0$ are occupied and 
all sites $x>0$ are empty at time 0 (Fig. 2).
In this case one can label the particles from the right
so that the particle $j$ ($j=1,2,\ldots$) starts from site $1-j$.
Several results for other initial and boundary conditions 
will be explained in sections 6 and 7. We denote the 
measure corresponding to the ASEP dynamics by $\mathbb{P}$.

The first step in this approach is to introduce 
the ``waiting time'' table, or the matrix $w=\{w_{i,j}\}$, 
in which the $(i,j)$ element $w_{i,j}$ is the waiting time 
of the $j$th particle before making the $i$th hop after the 
target site becomes empty. 
Let us see the example in Fig. 3.
The configuration does not change until the first particle 
moves to the site one. It took 1 step before the first move 
and hence $w_{1,1}=1$. Now two particles can move to the right. 
It took 1 step (resp. 2 steps) before the second (resp. first)
particle makes the first (resp. second) jump and hence
$w_{1,2}=1$ (resp. $w_{2,1}=2$). After the second jump of the first 
particle the second particle waits 2 steps before the next move
and hence $w_{2,2}=2$. One continues this process to get 
all $w_{i,j}$'s for a given sample of the dynamics of dTASEP.
For this example the matrix looks like
\begin{equation}
\label{wex}
 w=\begin{bmatrix}
    1 & 1 & 1 & 3 & \cdots \\
    2 & 2 & 1 & 0 & \\
    1 & 0 & 0 & 1 & \\
    1 & 2 & 1 & 0 & \\
    \vdots &&&& \ddots
   \end{bmatrix}.
\end{equation}
It is clear that the time evolution of the dTASEP is equivalent 
to the infinite matrix $w_{i,j}$.
If we call a matrix whose elements are nonnegative integers an 
$\mathbb{N}$ matrix, $w$ is an $\mathbb{N}$ matrix.
When considering various realizations of dynamics, 
each $w_{i,j}$ is a random variable and 
geometrically distributed with parameter $q$;
$ \mathbb{P}(w_{i,j}=l) = (1-q) q^l \,\, (l=0,1,2,...)$.

Next let us define $G(M,N)$ by
\begin{equation}
\label{Gdef}
 G(M,N) = \max_{\pi\in\prod_{M,N}}\{\sum_{(i,j)\in\pi} w_{i,j} \}.
\end{equation}
Here $\pi$ is a path from $(1,1)$ to $(M,N)$ which consists of 
elementary path connecting points $(i,j)\to(i+1,j)$ or 
$(i,j)\to(i,j+1)$. In the matrix representation as in (\ref{wex}),
$\pi$ is a down/right path, i.e., it starts from $(1,1)$ and goes 
either down or right until it reaches $(M,N)$. 
$\Pi_{M,N}$ is the collection of all such paths. 
If we regard $w_{i,j}$ as representing a value of a potential energy 
at a position $(i,j)$, (\ref{Gdef}) can be considered as the problem 
of the zero temperature directed polymer in random medium.
One can confirm oneself that $G(M,N)+M+N -1$ is equal to the 
time at which the $N$th particle has just made the $M$th step.
Considering the dynamics of the $N$th particle, one notices that 
$N(t) \geq N$ is equivalent to saying that the $N$th particle makes 
the $N$th jump before $t$. Hence we have
\begin{equation}
 \mathbb{P}[N(t) \geq N] = \mathbb{P}[G(N,N)+2N-1 \leq t].
\end{equation}
Though $w$ is in principle an infinite matrix, to study 
$N(t)$, we can restrict our attention to the $N\times N$
submatrix $\{w_{i,j}\}_{1\leq i,j \leq N}$ of $w$.
This is related to the Markov property and the fact that 
in the TASEP each particle can not affect the dynamics of 
the particles on its right. 
For instance, from the $3\times 3$ submatrix of $w$,
\begin{equation}
       \begin{bmatrix}
	1 & 1 & 2 \\
        2 & 2 & 1 \\
        1 & 0 & 0
       \end{bmatrix},
\label{w3}
\end{equation}
one sees that $G(3,3)=6$ with the maximizing path, 
$(1,1)\to(2,1)\to(2,2)\to(2,3)\to(3,3)$.
This gives $G(3,3)+5=11$ which agrees with 
the time at which the third particle 
has made the third hop in Fig. 3.

To proceed we use a bijection known as the 
Robinson-Schensted-Knuth(RSK) algorithm between an 
$\mathbb{N}$ matrix and a pair $(P,Q)$ of 
semistandard Young tableaux(SSYT). 
A short explanation about the SSYT and the Schur function 
is given in the appendix. For more details and the RSK algorithm, 
see for instance \cite{Stanley1999,Fulton1997}.
The pair of SSYTs corresponding to (\ref{w3}) is 
\begin{equation*}
 P=\young(111123,22,3), \quad  Q=\young(111222,22,3).
\end{equation*}
In the language of SSYT, the $G(N,N)$ is given by
\begin{equation}
 G(N,N) = \lambda_1
\end{equation}
where $\lambda_1$ is the length of the first row of $P$.
This is easily seen as a property of the RSK algorithm. 
For our example, $\lambda_1=6$ is the same as $G(3,3)$.
Therefore the problem of $N(t)$ is reduced to a combinatorial 
problem of SSYTs.
If we denote the number of SSYT of shape
$\lambda$ and entries from $\{1,2,\ldots,N\}$ by $L(\lambda,N)$, 
we have
\begin{align}
 \mathbb{P}[G(N,N)+2N-1 \leq t] 
 &=
 (1-q)^{N^2} \sum_{\lambda: \lambda_1 \leq t-2N+1} 
 q^{|\lambda|} L(\lambda,N)^2 \notag\\
 &=
 (1-q)^{N^2} \sum_{\lambda: \lambda_1 \leq t-2N+1} 
 q^{|\lambda|} s_{\lambda}(\underbrace{1,\ldots,1}_N,0,\ldots)^2 
\label{GL}
\end{align}
where a formula (\ref{Ls}) is used in the second equality.
Rewriting (\ref{GL}) using the formula (\ref{sp}) 
and $h_j=\lambda_1-j+1$, one obtains 
\begin{equation}
 \mathbb{P}[G(N,N)+2N-1 \leq t] 
 =
 \frac{1}{Z} 
 \sum_{h_j=-N+1}^{t-2N+1}
 \prod_{1\leq j<l\leq N} (h_j-h_l)^2
 \prod_{j=1}^N  q^{h_j}.
\label{NMeix}
\end{equation}
Here and in the following the symbol $Z$ is used to represent 
a normalization constant. For the present case it is 
\begin{equation}
 Z
 =
 \frac{q^{\frac12 N(N-1)}}{(1-q)^{N^2}}
 \prod_{j=1}^N j! (j-1)!.
\end{equation}
The expression of the RHS of (\ref{NMeix}) has a strong similarity
with a quantity appearing in the random matrix theory.
Hence generalizing and applying techniques from random 
matrix theory, one can study the asymptotic behavior of the 
quantity to obtain
\begin{equation}
 \lim_{N\to\infty}\mathbb{P}
 \left[\frac{G(N,N)-\frac{2\sqrt{q}}{1-\sqrt{q}}N}{dN^{1/3}} \leq s\right]
 =
 F_2(s)
\label{GF2}
\end{equation}
with $d=q^{1/6}(1+\sqrt{q})^{1/3}(1-\sqrt{q})^{-1}$.
In terms of $N(t)$, this is equivalent to 
\begin{equation}
 \lim_{t\to\infty}\mathbb{P}
 \left[ \frac{N(t)-\frac{1-\sqrt{q}}{2}t}{d' t^{1/3}} \geq -s \right]
 =
 F_2(s)
\label{NF2}
\end{equation}
with $d' = 2^{-4/3} q^{1/6} (1-q)^{1/3}$.
The function $F_2$ appearing on the RHS is called 
the GUE Tracy-Widom distribution function \cite{TW1994} and 
describes the distribution of the largest eigenvalue of the 
Gaussian unitary ensemble (GUE). 
Hence the above equation states that the current of the TASEP for 
the step initial condition is, in the appropriate scaling limit, 
equivalent to that of the largest eigenvalue of the GUE.
It is stressed that $F_2(s)$ is completely different from the 
error function which is the distribution function for the 
ordinary Gaussian fluctuation. 
This is the most basic and fundamental result in the approach of 
random matrix theory to the ASEP.   
Basics of random matrix techniques to obtain $F_2$ and related 
quantities will be given in the next section. 

Before ending the section, let us consider the cTASEP limit. 
In this limit, $w_{i,j}$ becomes an exponential random variable 
with parameter one and (\ref{NMeix}) reduces to 
\begin{equation}
\label{lue}
 \mathbb{P}[G(N,N)\leq t]
 =
 \frac{1}{Z} \int_{[0,t]^N}
 \prod_{1\leq j<l \leq N} (x_j-x_l)^2
 \prod_{j=1}^N e^{-x_j} d^N x
\end{equation}
with
\begin{equation}
 Z
 =
 \prod_{j=1}^N j! (j-1)!.
\end{equation}
The same quantity appears in the random matrix theory as the 
distribution of the largest eigenvalue in an ensemble called 
the Laguerre unitary ensemble. 
As a result, for the cTASEP, we have
\begin{equation}
 \lim_{t\to\infty} 
 \mathbb{P}\left[ \frac{N(t)-\frac{t}{4}}{2^{-4/3}t^{1/3}} \geq -s \right]
 = 
 F_2(s).
\end{equation}

\setcounter{equation}{0}
\section{Random matrix techniques}
In this section, we explain the basics of the techniques
from random matrix theory which would be helpful to understand 
the results for the ASEP. 
In physics the random matrices were first introduced to 
model a complex Hamiltonian of nuclei. 
But the random matrix theory has also found a lot of 
applications not only in nuclear physics but also in 
other branches of physics, mathematics, statistics and so on.
The basic reference of random matrix theory is \cite{Mehta2004}. 
The book by Forrester \cite{ForresterBook} is also very useful. 

A random matrix is a matrix with random elements. 
There are infinitely many variety of random matrix
ensembles but there are some random matrix ensembles 
with especially good properties. 
Among them are the Gaussian ensembles which are 
defined by the probability distribution of 
$N\times N$ matrices,
\begin{equation}
 P(H)dH = e^{-\frac{\beta}{2}\text{Tr} H^2} dH.
\label{RM}
\end{equation}
Here $\beta$ takes three different values $\beta=1,2,4$.
When $\beta=1$, $H$ is taken to be a real symmetric 
matrix and $dH = \prod_{j=1}^N dH_{jj} \prod_{j<l} dH_{jl}$.
Writing down the $\text{Tr}$ in (\ref{RM}) explicitly, one 
sees that this definition is equivalent 
to saying that each independent element obeys a Gaussian distribution
with variance 1 for the diagonal and 1/2 for the off-diagonal 
elements.  This ensemble is called the Gaussian orthogonal 
ensemble(GOE).
For $\beta=2$, $H$ is taken to be a Hermitian matrix 
and $dH = \prod_{j=1}^N dH_{jj} \prod_{j<l}
dH_{jl}^R dH_{jl}^I$ where $H_{jl}^R$ and $H_{jl}^I$
are the real and imaginary parts of $H_{jl}$ respectively. 
This is the Gaussian unitary ensemble (GUE). 
For $\beta=4$, the ensemble is called the Gaussian 
symplectic ensemble (GSE). For this case,
$H$ is taken to be a self-dual quaternion 
matrix, whose definition is not given here.

In many applications we are interested in the statistics 
of eigenvalues. For the Gaussian ensembles, we can derive 
the joint density function for eigenvalues explicitly.
Basically we change the coordinate from the original matrix 
elements to the eigenvalues and the diagonalizing matrix 
and then integrate out the latter degrees of freedom.  
The procedure is explained in \cite{Mehta2004,ForresterBook}
but is omitted here since it is not directly used in the 
applications to ASEP.
The resulting joint probability density function of 
eigenvalues $\{x_j\}_{j=1,\ldots,N}$ is 
\begin{equation}
 P_{N\beta}(x_1,\ldots,x_N) 
 =
 \frac{1}{Z}
 \prod_{1\leq j<l \leq N} |x_j-x_l|^{\beta}  
 \prod_{j=1}^N e^{-\beta x_j^2/2}
\label{PNb}
\end{equation}
for $\beta=1,2,4$.
Let $\mathbb{P}_{N\beta}$ denote the corresponding 
measure. Using this expression one can study in principle  
any statistical property of eigenvalues of the Gaussian ensembles. 
Here we focus on the statistical behavior of 
the largest eigenvalue.
From the joint eigenvalue distribution (\ref{PNb}), the probability 
that the largest eigenvalue $x_1$ is less than $u$ is given by
\begin{equation}
 \mathbb{P}_{N\beta}[x_1\leq u] 
 =
 \frac{1}{Z}\int_{(-\infty,u]^N} 
 \prod_{1\leq j<l\leq N} |x_j-x_l|^{\beta}
 \prod_{j=1}^N e^{-\beta x_j^2/2} d x_j.
\label{x1}
\end{equation}

Next for the case of $\beta=2$ we rewrite this expression 
to another which is more suitable for asymptotic analysis. 
The probability density (\ref{PNb}) for $\beta=2$ can be written as
\begin{equation}
 \frac{1}{Z}\det(\varphi_j(x_l))_{j,l=1}^N \det(\psi_j(x_l))_{j,l=1}^N 
\label{PN2}
\end{equation}
if we notice 
\begin{equation}
 \prod_{1\leq j<l\leq N} (x_l-x_j)
 =
 \det(x_j^{l-1})_{j,l=1}^N,
\end{equation}
and set $\varphi_j(x)=\psi_j(x) = x^{j-1} e^{-x^2/2}$.
We show the identity 
\begin{equation}
\frac{1}{Z}\int_{-\infty}^{\infty} dx 
\det(\varphi_j(x_l))_{j,l=1}^N \det(\psi_j(x_l))_{j,l=1}^N 
\prod_{j=1}^N (1+g(x_j)) 
=
\det(1+ K g). 
\label{detFdet}
\end{equation}
Here the determinant on the RHS is the Fredholm determinant 
defined as
\begin{equation}
\det(1+ K g) 
=
\sum_{m=0}^{\infty} \frac{(-1)^m}{m!}
\int_{-\infty}^{\infty}dx_1 \cdots \int_{-\infty}^{\infty}dx_m
g(x_1)\cdots g(x_m) \det(K(x_j,x_l))_{j,l=1}^m, 
\label{Fred}
\end{equation}
where the kernel is 
\begin{equation}
 K(x_1,x_2) 
 = 
 \sum_{j,l=1}^N \psi_j(x_1) [A^{-1}]_{j,l} \varphi_l(x_2)
\label{kernel}
\end{equation}
with the matrix $A$ being
\begin{equation}
 A_{j,l} = \int_{-\infty}^{\infty} dx \varphi_j(x) \psi_l(x).
\end{equation}
There are several ways of proof; here we follow \cite{TW1998}.
Using the Heine identity,
\begin{equation}
 \int dx_1 \cdots \int dx_N
 \det(\varphi_j(x_l))_{j,l=1}^N \det(\psi_j(x_l))_{j,l=1}^N 
 = 
 \det(\int dx \varphi_j(x) \psi_l(x))_{j,l=1}^N,
\end{equation}
the LHS of (\ref{detFdet}) can be rewritten as
\begin{align}
 &\quad 
 \det(A_{j,l}+\int dx \varphi_j(x) \psi_l(x) g(x)) / \det A 
 \notag\\
 &=
 \det (1_{jl}+\int dx (A^{-1}\varphi)_j(x)g(x)\cdot \psi_l(x)),
\end{align}
where 1 is the identity matrix of size $N$.
One can regard this as $\det(1+BC)$ if we set
\begin{equation}
 B_{j,x} = (A^{-1}\varphi)_j(x)g(x),\quad C_{x,j} = \psi_j(x) .
\end{equation}
Then using a simple identity $\det(1+B C) = \det(1+C B)$
and rewriting $B,C$ again in terms of $\varphi,\psi$,
we get (\ref{detFdet}). Some manipulations here are rather formal 
but can be justified \cite{TW1998}.
Since (\ref{x1}) with $\beta=2$ is written in the form 
of LHS of (\ref{detFdet}) with $g(x)= -\chi_{(u,\infty)}(x)$ where 
$\chi_{(u,\infty)}(x)=1(x>u),0(x\leq u)$, 
the probability that the largest eigenvalue $x_1$ is less than 
$u$ in the GUE is written in the form of the Fredholm determinant,
\begin{equation}
 \mathbb{P}_{N2}[x_1 \leq u] = \det(1+K g).
\end{equation}

To proceed, a nontrivial task is the computation of the 
inverse matrix $A^{-1}$.
For the GUE, this is (implicitly) done as follows. 
Let us remember that the addition of a row or a column to 
another does not change the value of the determinant.
Then the functions $\varphi_j$ and $\psi_j$ in (\ref{PN2}), 
which we have taken to be the monic polynomials $x^{j-1}$, can be
replaced by any polynomials of the same degree modulo a numerical factor. 
For the GUE, it is convenient to take 
\begin{equation}
 \varphi_j(x) = \psi_j(x) = e^{-x^2/2} H_j(x)
\end{equation}
where $H_n(x)$ is the Hermite polynomials \cite{AAR1999}, 
\begin{equation}
 H_n(x) = (-1)^n e^{x^2} \frac{d^n}{d x^n} e^{-x^2}.
\end{equation}
Using the orthogonality relation of the Hermite polynomials,
one sees that $A$ is diagonal,
\begin{equation}
 A_{j,l} = \int_{-\infty}^{\infty} H_j(x) H_l(x) e^{-x^2} dx 
         = \sqrt{\pi} 2^j j! \delta_{jl}.
\end{equation}
Then the kernel (\ref{kernel}) for the GUE, which we denote by 
$K_{N2}(x_1,x_2)$, is calculated to be 
\begin{equation}
 K_{N2}(x_1,x_2) 
 = 
 e^{-\frac12(x_1^2+x_2^2)}
 \sum_{n=0}^{N-1} \frac{H_n(x_1)H_n(x_2)}{\sqrt{\pi} 2^n n!} .
\end{equation}
This expression of the kernel allows one to study various 
statistical properties of the eigenvalues such as 
the eigenvalue density, the two point correlation function, etc.
To take a scaling limit for the largest eigenvalue,
we use the asymptotic expression of the Hermite polynomial
\cite{Szego1975},
\begin{equation}
 e^{-\frac12 x^2}H_n(x) 
 \sim 
 \pi^{\frac14}2^{(2n+1)/4} (n!)^{1/2} n^{-1/12} 
 \text{Ai}(2^{1/2}n^{1/6}(x-\sqrt{2n})),
\end{equation}
where $\text{Ai}(x)$ is the Airy function defined by 
\begin{equation}
 \text{Ai}(x) 
 = 
 \frac{1}{2\pi} \int_{-\infty+i\epsilon}^{\infty+i\epsilon} 
 e^{izx+\frac{i}{3} z^3} dz
\label{AiK}
\end{equation}
with $\epsilon>0$.
If we set $x_j = \sqrt{2N} + \xi_j/(\sqrt{2}N^{1/6})$ ($j=1,2$), 
one has for large $N$
\begin{equation}
 K_{N2}(x_1,x_2) \sim  \sqrt{2} N^{1/6} \K_2(\xi_1,\xi_2)
\end{equation}
where
\begin{equation}
 \K_2(\xi_1,\xi_2)
 =
 \frac{ \text{Ai}(\xi_1) \text{Ai}'(\xi_2)
  -\text{Ai}(\xi_2) \text{Ai}'(\xi_1)}{\xi_1-\xi_2} 
\end{equation}
is known as the Airy kernel.
Hence one obtains
\begin{equation}
 \lim_{N\to\infty} 
 \mathbb{P}_{N2}\left[ (x_1-\sqrt{2N})\sqrt{2}N^{1/6} \leq s \right] 
 =
 \det(1+\K_2 \mathcal{G})
 =:
 F_2(s).
\label{x1N2}
\end{equation}
The function $F_2(s)$ is defined as the Fredholm determinant 
(\ref{Fred}) with the kernel $\K_2$ (\ref{AiK})
and $\mathcal{G}(x) = -\chi_{(s,\infty)}$.
The function $F_2(s)$ describes the fluctuation of the 
largest eigenvalue of GUE in the scaling limit and is
called the GUE Tracy-Widom distribution function.
It is known to have a different representation 
using the solution to the Painlev\'e II equation\cite{TW1994}. 
Equation (\ref{x1N2}) says that the largest eigenvalue $x_1$
in the GUE behaves like $x_1\sim \sqrt{2N}$ in average, has a fluctuation 
of order $O(N^{-1/6})$ and has the $F_2$ as the  
distribution function in the $N\to\infty$ limit if appropriately scaled.

The function $F_2(s)$ is the same as the one which appeared in 
the RHS of (\ref{GF2}) and (\ref{NF2}).
Notice a close similarity between (\ref{NMeix}) and (\ref{x1})
with $\beta=2$.
The main difference in (\ref{NMeix}) is that the Gaussian 
$e^{-x^2}$ ($x\in\mathbb{R}$) is replaced by $q^h$ ($h=0,1,2,\ldots$).
As a consequence we use the orthogonal polynomials which are 
orthogonal with respect to the weight $q^h$, which is known as
the Meixner polynomials \cite{KS1996p}. 
Then (\ref{NMeix}) could be called the 
Meixner ensemble representation for the current of the TASEP. 
The derivation of (\ref{GF2}) from (\ref{NMeix}) is more involved 
but is similar to the one for the GUE \cite{Johansson2000}.

For GOE($\beta=1$) and GSE($\beta=4$) 
the power of the determinant in (\ref{x1}) is different from 2
and the above calculation does not work.
But it is possible to modify the arguments and obtain analogous results
\cite{TW1996,TW1998}. They read
\begin{equation}
 \lim_{N\to\infty}\mathbb{P}_{N\b}
 \left[(x_1-\sqrt{2N})\sqrt{2}N^{1/6} \leq s \right]
 =
 F_{\b}(s)
\label{x1Nb}
\end{equation}
with $\beta=1,4$. The functions $F_1(s)$ (resp. $F_4(s)$) is called 
the GOE (resp. GSE) Tracy-Widom distribution function. 
They are expressed as Fredholm determinants with $2\times 2$ matrix kernel. 
The Painlev\'e II representation is also known \cite{TW1996}. 

A remark is in order. 
To compare the results of ASEP with those of the Gaussian ensembles, 
it is sometimes more convenient to change the normalization from 
(\ref{RM}) to 
\begin{equation}
 P(H)dH = e^{-\frac{\beta}{4N} \text{Tr} H^2}dH.
\end{equation}
Then (\ref{x1N2}) and (\ref{x1Nb}) are replaced by
\begin{equation}
 \lim_{N\to\infty}\mathbb{P}_{N\b}\left[\frac{x_1-2N}{N^{1/3}} \leq s\right]
 =
 F_{\b}(s)
\label{F14}
\end{equation}
for $\beta=1,2,4$ in which the exponents of $N$ are the same as in 
(\ref{GF2}).

Let us now consider a time dependent version of the random 
matrix ensemble known as the Dyson's Brownian motion
\cite{Dyson1962}.
The random matrix $H(t)$ now depends on time $t$.
Each independent component of the matrix performs 
the Ornstein-Uhlenbeck process 
for which the Fokker-Plank equation for the probability density 
$p(t,x)$ takes the form,
\begin{equation}
 \frac{\partial}{\partial t} p(t,x)
 =
 D \frac{\partial^2}{\partial x^2}p(t,x) 
 +
 \frac{\partial}{\partial x} (xp(t,x)).
\end{equation}
Here $D$ is the diffusion constant and should be taken appropriately
depending on the ensembles in question.
Let us consider the stochastic dynamics of the eigenvalues
of this random matrix. We are interested in the multi-time 
joint distribution of the eigenvalues.
It is known that, for general $\beta$, time evolution of the 
probability density of the 
eigenvalues is described by an equation which can be mapped 
to the quantum Calogero model \cite{ForresterBook}. 
For the GUE($\beta=2$) case, the system is a free fermion and 
we can get detailed information from this fact as we see below. 
For the GOE($\beta=1$) and the GSE($\beta=4$), however, 
the analysis becomes much more involved and the computation of 
correlation functions have so far been very restricted.

Let us consider the stationary situation for the GUE in which 
the dynamics and the initial conditions are both taken to be 
the GUE type. We call this the tGUE (This is {\it not} a standard
notation). If we denote the eigenvalues of the matrix $H_{j,l}(t)$ by 
$x_j^t$  ($j=1,2,\cdots,N$), the dynamics of them look like
Fig. 4. The joint density function at two times can be 
written as a product of determinants,
\begin{equation}
 P(x_1^{t_1},\ldots,x_N^{t_1},x_1^{t_2},\ldots,x_N^{t_2}) 
 =
 \frac{1}{Z}
 \det(\varphi^{(t_1)}_j(x_l^{t_1}))_{j,l=1}^N 
 \cdot 
 \det(\phi_{t_1,t_2}(x_j^{t_1},x_l^{t_2}))_{j,l=1}^N \cdot
 \det(\psi^{(t_2)}_j(x_l))_{j,l=1}^N
\label{GUE2t}
\end{equation}
where 
$\varphi^{(t_1)}_j(x) = \psi^{(t_2)}_j(x) = e^{-x^2/2} x^{j-1}$ 
($j=1,2,\ldots,N$) and
\begin{equation}
 \phi_{t_1,t_2}(x_1,x_2) 
 = 
 \frac{1}{\sqrt{(1-e^{-2(t_2-t_1)})\pi}}
 \exp\left[-\frac{(x_2-e^{-(t_2-t_1)}x_1)^2}{1-e^{-2(t_2-t_1)}}\right].
\end{equation}
This is a generalization of (\ref{PN2}) to the two time case. 
The derivation can be found in \cite{Mehta2004,ForresterBook} but
is omitted here since it is again not directly used in the applications 
to ASEP. Furthermore the joint probability that the largest eigenvalue
$x_1^{t_j}$ (the height of the top curve at $t_j$ in Fig. 4)
is less than $u_j$ at time $t_j$ ($j=1,2$) 
is again given by a Fredholm determinant.
Let us introduce a matrix
\begin{equation}
 A_{jl} = 
 \int_{-\infty}^{\infty} dx_1 \int_{-\infty}^{\infty} dx_2 
 \varphi^{(t_1)}_j(x_1) \phi_{t_1,t_2}(x_1,x_2) 
 \psi_l^{(t_2)}(x_2),
\label{AtGUE}
\end{equation}
and 
\begin{align}
 \varphi_j^{(t_2)}(x_2) 
 &= 
 \int_{-\infty}^{\infty} dx_1 \varphi_j^{(t_1)}(x_1) 
 \phi_{t_1,t_2}(x_1,x_2), \\
 \psi_j^{(t_1)}(x_1) 
 &= 
 \int_{-\infty}^{\infty} dx_2 ~ \phi_{t_1,t_2}(x_1,x_2) 
 \psi_j^{(t_2)}(x_2).
\end{align}
Then we have 
\begin{equation}
 \mathbb{P}_{N2}[x_1^{t_1}\leq u_1,x_1^{t_2}\leq u_2]
 =
 \det(1+K_{N2} g)
\label{x1Nt}
\end{equation}
where the kernel is 
\begin{align}
 K_{N2}(t_1,x_1;t_2,x_2)
 &=
 \tilde{K}_{N2}(t_1,x_1;t_2,x_2) - \phi_{t_1,t_2}(x_1,x_2), \\
 \tilde{K}_{N2}(t_1,x_1;t_2,x_2)
 &= 
 \sum_{j,l=1}^N \psi_j^{(t_1)}(x_1) [A^{-1}]_{jl} \varphi_l^{(t_2)}(x_2),\\
 g(t_j;x) &= -\chi_{(u_j,\infty)}(x), \quad (j=1,2), 
\end{align}
with the convention that $\phi_{t_1,t_2}(x_1,x_2)=0$ when
$t_1 \geq t_2$. 
In this kernel $(t_1,t_2)$ should be understood as 
a representative of either $(t_1,t_1),(t_1,t_2),(t_2,t_1),(t_2,t_2)$.
Similar abuse of notation will be used in rest of the paper as well.
The formula (\ref{x1Nt}) can be derived by a generalization of the 
argument to get (\ref{detFdet}).

For the tGUE, the matrix $A$ in (\ref{AtGUE}) is again diagonalized 
by taking $\varphi_j^{(t_1)}$ and $\psi_j^{(t_2)}$ to be 
the Hermite polynomials. 
Then the kernel turns out to be 
\begin{equation}
 \tilde{K}_{N2}(t_1,x_1;t_2,x_2) 
 =
 e^{-\frac12(x_1^2+x_2^2)}\sum_{n=0}^{N-1} 
 \frac{H_n(x_1)H_n(x_2)}{\sqrt{\pi} 2^n n!} 
 e^{-(n+1/2)(t_2-t_1)}.
\end{equation}
Using the integral formulas for the Hermite polynomial
\cite{AAR1999},
we can rewrite this in a double integral form,
\begin{equation}
 \tilde{K}_{N2}(t_1,x_1;t_2,x_2) 
 =
 e^{-\frac12(x^2-y^2)} \frac{\sqrt{2} e^{(N+\frac12)(t_1-t_2)}}{(2\pi i)^2} 
 \int_{\gamma} dz_1 \int_{\Gamma} dz_2 \frac{z_1^N}{z_2^N}
 \frac{e^{ -\frac{z_1^2}{2}+\sqrt{2}x_1 z_1 + \frac{z_2^2}{2}-\sqrt{2}x_2 z_2}}
      {z_1e^{t_1-t_2}-z_2} 
\label{KN2t}
\end{equation}
where $\Gamma$ is a contour enclosing the origin anticlockwise
and $\gamma$ is an any path from $-i\infty$ to $i\infty$.
This representation of the kernel has a form suitable for the 
saddle point method. Performing the asymptotic analysis, one gets 
with $t_j=\frac{\tau_j}{N^{1/3}}$
\begin{equation}
 \lim_{N\to\infty}\mathbb{P}_{N2}
 \left[\left( x_1^{t_j}-\sqrt{2N}\right) 
              \sqrt{2}N^{1/6} \leq s_j \,\, (j=1,2) \right]
 =
 \det(1+\K_2 \mathcal{G}).
\end{equation}
Here $\mathcal{G}(\tau_j;x) = -\chi_{(s_j,\infty)}(x)$ ($j=1,2$)
and the kernel is 
\begin{equation}
\label{eAiK}
 \K_2(\t_1,\xi_1;\t_2,\xi_2)
 =
 \tilde{\K}_2(\t_1,\xi_1;\t_2,\xi_2)
 -
 \Phi_2(\t_1,\xi_1;\t_2,\xi_2)
\end{equation}
where
\begin{align}
 \tilde{\K}_2(\t_1,\xi_1;\t_2,\xi_2)
 &=
 \int_0^{\infty} d\lambda e^{-\lambda(\t_1-\t_2)} \Ai(\xi_1+\lambda) \Ai(\xi_2+\lambda), \\
 \Phi_2(\t_1,\xi_1;\t_2,\xi_2)
 &=
 \begin{cases}
 \int_{-\infty}^{\infty} d\lambda e^{-\lambda(\t_1-\t_2)} \Ai(\xi_1+\lambda) \Ai(\xi_2+\lambda),   
 & \t_1<\t_2, \\
 0, & \t_1 \geq \t_2.
 \end{cases}
\end{align}
This is a generalization of (\ref{x1N2}).
This describes the stochastic dynamics of the largest eigenvalue 
of the tGUE in the scaling limit and is called the 
Airy process 
\cite{PS2002b,Johansson2003}.

The weight (\ref{GUE2t}) is for the stationary situation. 
One can also consider the case where the dynamics is still 
$\beta=2$ but the initial condition is taken to be GOE or GSE. 
This describes the transition from the GOE or GSE to the GUE. 
For these transition ensembles, a similar analysis as above 
is possible \cite{FNH1999}.
For the dynamics of the GOE($\beta=1$) and GSE($\beta=4$) cases, 
however, much less is known as already mentioned.

\setcounter{equation}{0}
\section{Discrete polynuclear growth model}
In section 3, we have obtained a formula for the current 
fluctuation by mapping the problem to that of combinatorics 
of Young tableaux. In this section we derive a generalized 
result by introducing the discrete polynuclear growth 
(dPNG) model \cite{Johansson2003}.

The dPNG model can be defined by interpreting $G(i,j)$
as a height of a surface. Let us introduce a new coordinate 
(Fig. 5), 
\begin{equation}
 s=i+j-1, \quad x=i-j
\end{equation}
and set $h^x(s)=G(i,j)$. 
In the dPNG model $s$ and $x$ play the role of time and 
space respectively and 
$h^x(s)$ is considered as the height of a surface
at time $s$ and at position $x$.
From the definition of $G(M,N)$ (\ref{Gdef}), 
the growth dynamics of $h^x(s)$ is given by 
\begin{equation}
 h^x(s+1) = \max\{h^{x-1}(s),h^{x+1}(s)\}
           +w_{\frac{s+x}{2}+1,\frac{s-x}{2}+1}.
\end{equation}
Here $w_{i,j}$ is the waiting time matrix $w=\{w_{i,j}\}$ and 
is taken to be zero if $(s+x)/2+1$ or $(s-x)/2+1$
is not an integer. 
The time evolution of $h^x(s)$ until just before $s=3$ for our 
example is depicted in Fig. 6. 

The above definition can be restated as follows. 
At time $s=0$, the surface is flat; $h^x(0) = 0$ 
for all $x$. 
At each time $s$, nucleations occur at sites 
$x=-s+1,-s+3,\ldots,s-3,s-1$. A nucleation at each site 
is independent and the height of a nucleation at each site 
obeys the geometric distribution with parameter $q$.
In the meantime the surface also grow laterally 
in both directions deterministically with unit speed. 
Sometimes two steps collide whence the height is taken 
to be the higher one. The dPNG model is a model of 
stochastically growing surface. 

We further introduce the multi-layer version of the 
dPNG model as follows.
Suppose that there are infinitely many layers below 
the original surface. We denote the height of the $j$th layer 
by $h_j^x(s)$ ($j=1,2,\ldots$). In particular $h_1^x(s) = h^x(s)$.
Initially they are equally spaced as $h_j^x(0)=1-j$.
The growth of the $j$th ($j \geq 2$) layer is 
determined by the layer above. When there occurs a 
collision with height $l$ at $(j-1)$th layer, 
then there occurs a nucleation with height $l$ at 
the same position of the $j$th layer. They also 
grows laterally in both directions with unit speed 
but there are no stochastic nucleations. 
This is the multi-layer PNG model. A snapshot of a 
Monte-Carlo simulation is shown in Fig. 7. 

For a given time $s$ if we regard the PNG space 
coordinate  $x$ as time and the height coordinate 
$h$ as space, the multilayers can be considered as 
a system of random walkers with the non-colliding condition.
This kind of object has been known as vicious walks
in statistical physics \cite{Fi1984}. 
The weight of the non-colliding walkers (or, line 
ensembles) is known to be simply given by a product 
of determinants with entries being transition weight 
of each walker. 
Mathematically this follows from the 
Lindstr\"om-Gessel-Viennot theorem \cite{Lin1973,GV1989}.
Since our main interest is to study $N(t)$ and 
$h^0(2N-1)=G(N,N)$, let us fix $s=2N-1$ and 
write $h_j^x=h_j^x(2N-1)$ in the sequel.
The transition weight of each walker at each time step is 
given by
\begin{equation}
 \phi^{\text{dP}}_{x,x+1}(h_1,h_2) 
 =
 \begin{cases}
  (1-\sqrt{q})q^{(h_2-h_1)/2}, & h_2\geq h_1, \\
  0,                      & h_2<h_1, 
 \end{cases} \\
\end{equation}
when $x$ is odd and 
\begin{equation}
 \phi^{\text{dP}}_{x,x+1}(h_1,h_2) 
 =
 \begin{cases}
  0,                      & h_2>h_1, \\
  (1-\sqrt{q})q^{(h_1-h_2)/2}, & h_2\leq h_1,
 \end{cases} \\
\end{equation}
when $x$ is even. This is a consequence of the fact 
that a nucleation of a height $l$ occurs with probability 
proportional to $q^l$.
The transition weight from $x_1$ to $x_2$ is 
given by the convolution from $\phi_{x_1,x_1+1}$ through
$\phi_{x_2-1,x_2}$ and is compactly given by an integral 
formula,
\begin{equation}
 \phi_{x_1,x_2}^{\text{dP}}(h_1,h_2)
 =
 \begin{cases}
 \frac{(1-\sqrt{q})^{x_2-x_1}}{2\pi i}
 \int_{C_{1}} dz z^{h_2-h_1-1}
 (1-\sqrt{q}z_1)^{(x_1-x_2)/2} (1-\sqrt{q}/z_1)^{(x_1-x_2)/2},
 & x_1 < x_2, \\
 0, & x_1 \geq x_2,
 \end{cases}
\end{equation}
where $C_R$ is the contour enclosing the origin anticlockwise
with radius $R$.
The weight of the vicious walks that their positions 
are $\{h_j^x\}$ is simply given by a 
product of determinants of this transition weight,
\begin{equation}
\label{w_dPNG}
 \prod_{x=-2N+1}^{2N-2} \det(\phi^{\text{dP}}_{x,x+1}(h_j^x,h_l^{x+1}))
\end{equation}
with the initial and final condition  
$h_j^{-2N+1} = h_j^{2N-1} = 1-j$ ($j=1,2,\ldots$). 
Since the number of layers is infinite, the determinant on 
RHS is in principle infinite dimensional. 
But it can be regularized by restricting the number of layers
to be finite for a moment and then taking the limit. 

Notice a similarity of this weight (\ref{w_dPNG}) to the 
joint density of tGUE (\ref{GUE2t}).
One can use a similar argument to express the joint distribution 
of the top layer in the form of the Fredholm determinant. 
It reads
\begin{equation}
 \mathbb{P}[h_1^{x_1} \leq u_1,h_1^{x_2} \leq u_2] 
 = 
 \det(1+K_N^{\text{dP}}g)
\end{equation}
where the kernel is (for even $x_1,x_2$)
\begin{equation}
 K_N^{\text{dP}}(x_1,h_1;x_2,h_2) 
 =
 \tilde{K}_N^{\text{dP}}(x_1,h_1;x_2,h_2) 
 - 
 \phi_{x_1,x_2}^{\text{dP}}(h_1,h_2)
\end{equation}
with 
\begin{equation}
 \tilde{K}_N^{\text{dP}}(x_1,h_1;x_2,h_2)
 =
 \sum_{j,l=1}^{\infty} \psi_j^{(x_1)}(h_1) [A^{-1}]_{j,l} 
                       \varphi_l^{(x_2)}(h_2).
\end{equation}
Here 
\begin{equation}
 A_{j,l} = \phi^{\text{dP}}_{-2N+1,2N-1}(1-j,1-l),
\end{equation}
\begin{align}
 \varphi_{j-1}^{(x)}(h) &= \phi^{\text{dP}}_{-2N+1,x}(1-j,h), \\
 \psi_{j-1}^{(x)}(h)    &= \phi^{\text{dP}}_{x,2N-1}(h,1-j),
\end{align}
for $j,l=1,2,\ldots$.
The next step is the computation of the inverse matrix $A^{-1}$.
For the present case, it can be obtained by utilizing the fact 
that $A$ is a Toeplitz matrix, leading to the kernel\cite{Johansson2000}, 
\begin{equation}
 \tilde{K}_N^{\text{dP}}(x_1,h_1;x_2,h_2) 
 =
 \frac{(1-\sqrt{q})^{x_2-x_1}}{(2\pi i)^2}
 \int_{C_{R_1}} \frac{dz_1}{z_1^{h_1}} \int_{C_{R_2}} dz_2 
 \frac{z_2^{h_2-1}}{z_1-z_2} 
 \frac{(1-\sqrt{q}/z_1)^{N+x_1/2} (1-\sqrt{q}z_2)^{N-x_2/2}}
      {(1-\sqrt{q}z_1)^{N-x_1/2} (1-\sqrt{q}/z_2)^{N+x_2/2}},
\end{equation}
where $\sqrt{q} < R_2 < R_1<1/\sqrt{q}$
\cite{Johansson2003}.
This is similar to the kernel for the tGUE (\ref{KN2t}). 

Our original interest was to consider $h^0(2N-1)$
which is directly related to $N(t)$.
But in the surface growth picture it is natural to extend 
our analysis to the equal-time (in $s$) multi-point 
distribution because it is related to the roughness of 
the surface. One can show that the scaled height 
\begin{equation}
 \frac{
 h^x(s=2N-1)-a(b_0 + \frac{c(b_0)\tau}{N^{1/3}})N}
 {d(b_0) N^{1/3}}
\label{sc_h}
\end{equation}
with $x=2b_0 N + 2c(b_0)N^{2/3}\tau$
tends to the Airy process as $N\to\infty$ where 
\begin{align}
\label{defa}
 a(b) &= \frac{2\sqrt{q}}{1-q} \left( \sqrt{q}+\sqrt{1-b^2} \right),\\
 d(b) &= \frac{q^{\frac16}}{(1-q)(1-b^2)^{\frac16}}
             (\sqrt{1+b}+\sqrt{q}\sqrt{1-b})^{\frac23}
             (\sqrt{1-b}+\sqrt{q}\sqrt{1+b})^{\frac23}, \\
 c(b) &= q^{-\frac16} (1-b^2)^{\frac23}
             (\sqrt{1+b}+\sqrt{q}\sqrt{1-b})^{\frac13}
             (\sqrt{1-b}+\sqrt{q}\sqrt{1+b})^{\frac13}.
\end{align}
For $b_0=0$ this was proved in \cite{Johansson2003};
the general $b$ case is obtained as a special case 
of the results in \cite{IS2004}. 
Also notice that if we take $x=0$ and consider the 
fluctuation at the origin, this reduces to (\ref{GF2}).
It is possible to translate this result into the language of 
the TASEP by noticing that $h^x(s)$ corresponds to the time 
at which the $\frac{s-x+1}2$th particle has made the 
$\frac{s+x+1}2$th hop. 
It gives a joint distribution of the times 
at which ASEP particles pass certain bonds. 
It should be remarked here that the equal-time 
in PNG picture does not correspond to the equal-time 
in TASEP. 

In our argument the multi-layers were introduced as an auxiliary
object which makes it possible to analyze the properties of the 
original surface related to the quantities in TASEP. 
From the random matrix point of view, the multi-layers have a meaning
that in the limit the behaviors of the $j$th layer is the same as 
the $j$th largest eigenvalue of GUE. 
Moreover if we set 
\begin{equation}
 h_j^x = \lambda_j^x-j+1 \quad (j=1,2,\ldots),
\label{h_lam}
\end{equation}
the probability distribution of $\{\lambda_j^x\}$ is known to be a
special case of the Schur process \cite{OkoRe2003,BR2006}
which assigns the weight of the form,
\begin{equation}
s_{\lambda^{(1)}}(\rho_0^+)
s_{\lambda^{(1)}/\mu^{(1)}}(\rho_1^-)s_{\lambda^{(2)}/\mu^{(1)}}(\rho_1^+)
\cdots
s_{\lambda^{(N)}/\mu^{(N-1)}}(\rho_{N-1}^+)s_{\lambda^{(N)}}(\rho_N^-),
\end{equation}
on the set of Young diagrams satisfying
\begin{equation}
\lambda^{(1)}\supset\mu^{(1)}\subset\lambda^{(2)}\supset\mu^{(2)}
\cdots\supset\mu^{(N-1)}\subset\lambda^{(N)}.
\end{equation}
Here $s_{\rho/\mu}(\rho)$ is the Schur function with the specialization 
of algebra $\rho$. 
In particular for $x=0$, 
the probability distribution of $\lambda=\{\lambda_j^0\}_{j=1,2,\ldots}$ is 
proportional to 
\begin{equation}
 s_{\lambda}(\underbrace{\sqrt{q},\ldots,\sqrt{q}}_N,0,\ldots)^2.
\end{equation}
Restriction of our attention only to $\lambda_1$ 
gives another derivation of (\ref{GL}).

The limiting case of cTASEP has been studied in detail in 
\cite{Spohn2005,FerrariSpohn2006}.

In the PNG picture, there is another limit of much interest: 
$q\to 0,N\to\infty$ with $\sqrt{q}N$ fixed. This corresponds 
to the continuous PNG model in which time $s$ and space $x$
coordinate are continuous and the height $h$ takes integer values. 
In the study of the 1D KPZ systems using random matrix techniques,
this model has been studied most extensively 
\cite{PS2000a,PS2000b,PS2002a,PS2002b,PS2002p}.
From the universality there is little doubt that the continuous 
PNG model and the TASEP give the same results for the quantities 
which are not model dependent and characteristic of the KPZ 
universality class.

\setcounter{equation}{0}
\section{Variants}
So far we have concentrated on the current fluctuation 
for the step initial condition. It is also possible to 
generalize the analysis to some other initial conditions, 
boundary conditions and quantities.

\subsection{External sources}
One can generalize initial conditions to the $\rho_-, \rho_+$ 
situation in which the sites on the negative side are 
occupied with the stationary measure with density $\rho_-$
and the sites on the positive side with the one with density 
$\rho_+$. Of course $\rho_-=1,\rho_+=0$ corresponds to the 
step initial condition. 
In the stationary measure with density $\rho$ 
the waiting time of each particle is geometrically 
distributed with parameter $1-J(\rho)/\rho$. In the language
of $G(i,j)$ this situation can be realized by setting 
$G(1,j)$ (resp. $G(i,1)$) as a geometric random variable 
with parameter $1-J(\rho_-)/\rho_-$ (resp. $1-J(\rho_+)/\rho_+$).
This interpretation is explained in more detail for the 
cTASEP case in \cite{PS2002a}. In the dPNG model this corresponds 
to the model with external sources. The fluctuation of the height at the 
origin was studied in \cite{BR2000} and those at multi-points 
in \cite{IS2005}. They are readily translated into the 
language of the ASEP. 

\subsection{Stationary Two point function}
One of the most important achievements in the application of 
the methods of random matrix theory is that the stationary 
two point function for the 1D KPZ universality class has 
been computed quite explicitly by Ferrari, Pr\"ahofer and Spohn. 
In the PNG picture this is related to the limiting case of 
$\rho_- = \rho_+$ for the model with external sources above.
We do not go into details here. 
See the references \cite{PS2002a,PS2002p,FerrariSpohn2006}. 

\subsection{Alternating initial condition.}
Another tractable initial condition is the
alternating initial condition in which initially
all even sites are occupied and all odd sites are empty
(Fig. 8). 
In the dPNG picture the surface is initially flat.
The nucleations occur at all even sites at even time
and at all odd sites at odd time. This should be 
contrasted to the step case in which the nucleations at time 
$s$ occur only at sites $|x|<s$. 
In addition although the alternating case macroscopically looks 
the same as for the stationary case with density $\rho=1/2$, it 
turns out that the fluctuation is different. 

The fluctuation of $G(N,N)$ can be studied by 
using a symmetry property of the combinatorial 
approach and is shown to be $F_1(s)$
which appeared in (\ref{F14}) as the largest eigenvalue 
distribution of the GOE
\cite{BR2001b,BR2001c}.
It is also possible to define the multi-layer PNG model.
From the symmetry, the weight of $\lambda$ defined from 
the height of multilayers by (\ref{h_lam}) is given by
\begin{equation}
 s_{\lambda}(\underbrace{\sqrt{q},\ldots,\sqrt{q}}_N,0,\ldots)
\end{equation}
where partition $\lambda$ is restricted to ones such that 
it has the form $\lambda=2\mu$ where $\mu$ is another partition. 
From this one can show that 
the fluctuations of the multi-layers at the origin are the same 
as the largest eigenvalues of GOE. Combined with 
the fact that the flat case is statistically translationaly
invariant, this suggests that the multi-layers are the same 
as the GOE Dyson's Brownian motion (tGOE) as conjectured 
in \cite{Ferrari2004}.
But unfortunately the analysis of the multi-layers for the 
flat case has been difficult. 
For the multi point joint distribution, at least for now,
we need to employ a little different method
which will be explained in subsection 7.4.  

\subsection{Half-infinite system}
In the study of the ASEP, the boundary effects are important. 
The random matrix techniques also allow to study 
some properties of the half-infinite system in which the 
particle can enter the system at the origin with probability 
$\gamma \sqrt{q}$ ($0<\gamma< 1/\sqrt{q}$) if the site is empty. 
In the language of $G(i,j)$ this can be realized by putting the 
symmetry condition $G(i,j) = G(j,i)$. 
The symmetrized version of combinatorial approach was studied 
in \cite{BR2001b,BR2001c}. As the fluctuation of $G(N,N)$
there appear $F_1(s)$ and $F_4(s)$ in (\ref{F14}) depending 
on the value of $\gamma$.
In the ASEP language this says that the fluctuation of the 
current at the origin, i.e., the number of particles which entered 
the system in time $t$ is equivalent to that of the largest 
eigenvalue of the GOE and GSE in the scaling limit. 
The current at other points were studied in \cite{SI2004}. 
There the fluctuation of the current near the origin is shown to 
be described by the transition ensembles mentioned in section 3. 

\subsection{Position of particles}
So far we have mainly studied the fluctuation of the 
current, i.e. the number of particles which crossed 
a given bond until time $t$. It is also interesting 
to consider a related quantity, the position of 
particles at a given time $t$. 
In this section, we consider the step initial condition.

Let us denote the position of the $N$th particle at time $t$
by $x_N(t)$.
If we remember that $G(M,N)+M+N-1$ is the time at which 
the $N$th particle has made the $M$th jump, 
properties of $x_N(t)$ for a single particle can be 
obtained by a translation of results for the current. 
If we take $t\to\infty$ limit for a fixed $N$, it is 
known that $x_N(t)$ goes to the dynamics of the largest 
eigenvalue of GUE \cite{IS2007p}. 
Here we consider another limit 
$t\to\infty, N\to\infty$ with $\nu=N/t$ fixed. 
The quantity $x_{\nu t}(t)/t$ has a deterministic limit,
\begin{equation}
 \lim_{t\to\infty} (x_{\nu t}(t)+\nu t-1)/t 
 =
 a^*(\nu)
\end{equation}
where
\begin{equation}
 a^*(\nu) 
 = 
 \begin{cases}
   1-q -(1-2q)\nu -2\sqrt{\nu(1-\nu)q(1-q)}, & 0<\nu<1, \\ 
   0, & \nu>1.
 \end{cases}
\end{equation}
For a single particle, the fluctuation of $x_N(t)$ 
around this position can also be obtained from the result 
for the current by setting $\tau=0$ in (\ref{sc_h}).

It is further possible to study the equal time joint 
distribution of the position of several particles by 
introducing a multi-layer PNG model which is a little 
different from the one in section 5. See Fig. 9.
It is defined as 
follows. Let $k_j^v(t)$ be a height of the $j$th 
($j=1,2,\ldots$) layer at time $t$ ($=0,1,2,\ldots$)
and at position $v=0,1,2,\ldots$. Initially they 
are all flat; $k_j^v(0)=1-j$ for all $v$. At each 
time $t$ at each layer nucleations of height one may 
occur at sites $v=1,3,\ldots,2t-1$ with the condition 
that the layers do not touch. Here a nucleation at 
each site is independent, has a shape depicted as in 
Fig. 9,  and occurs with probability $q$. 
In the meantime it grows laterally to the right with 
speed 2. 
Notice a similarity and difference between the PNG 
model here and that in section 5. 
This is the same as the DR-paths in the Aztec 
diamond \cite{Johansson2002, Johansson2005}. 
One can confirm oneself that the dynamics of the 
top layer $k^v(t) = k_1^v(t)$ gives, if set
\begin{equation}
x_N(t) = t-2N+2-k^{2N-1}(t),
\label{xk}
\end{equation}
the correct statistical properties of the TASEP particle positions. 

The multi-layers in this PNG model can again be regarded as 
vicious walks.
The transition weight of each walker is now
\begin{equation}
 \phi^*_{v,v+1}(k_1,k_2) 
 =
 \begin{cases}
  p^{k_2-k_1}, & k_2\geq k_1, \\
  0,           & k_2 < k_1, 
 \end{cases} \\
\end{equation}
with $p=q/(1-q)$ when $v$ is even and 
\begin{equation}
 \phi^*_{v,v+1}(k_1,k_2) 
 =
 \begin{cases}
  1, & k_2=k_1,k_1-1, \\
  0, & \text{otherwise},
 \end{cases} \\
\end{equation}
when $v$ is odd.
The weight of the vicious walks are simply given by a 
product of determinant of this transition weight,
\begin{equation}
\label{w_dPNGd}
 \prod_{v=0}^{2N-1} \det(\phi^*_{v,v+1}(k_j^v,k_l^{v+1}))
\end{equation}
with the initial and final condition  
$k_j^0 = k_j^{2N} = 1-j$ ($j=1,2,\ldots$). 

The joint distribution has the form,
\begin{align}
 &\quad
 \mathbb{P}[x_{r_1}(t) \geq t-2r_1+2-u_1,
            x_{r_2}(t) \geq t-2r_2+2-u_2] \notag\\
 &=
 \mathbb{P}[k^{2r_1-1}(t) \leq u_1, k^{2r_2-1}(t) \leq u_2] 
 =
 \det(1+K_t^* g)
\end{align}
where the kernel is given by
\begin{equation}
 K_t^*(r_1,k_1;r_2,k_2) 
 =
 \tilde{K}_t^*(r_1,k_1;r_2,k_2) 
 -
 \phi^*_{r_1,r_2}(k_1,k_2),
\end{equation}
with
\begin{align}
 K_t^*(r_1,k_1;r_2,k_2) 
 &= 
 \frac{1}{(2\pi i)^2}
 \int_{C_{R_1}} \int_{C_{R_2}} dz_1 dz_2 \frac{z_1^{k_1-1}}{z_2^{k_2+1}}
 \frac{z_2}{z_2-z_1} 
 \frac{(1+1/z_1)^{t-r_1+1}(1-pz_1)^{r_1}}
      {(1+1/z_2)^{t-r_2+1}(1-pz_2)^{r_2}}, \\
 \phi^*_{r_1,r_2}(k_1,k_2)
 &=
 \frac{1}{2\pi i}\int_{C_R} dz 
 \frac{1}{z^{k_2-k_1+1}} \left( \frac{1+1/z}{1-pz} \right)^{r_2-r_1}
\end{align}
where $R_1 < R_2$ and $0<R<1$.
Applying the asymptotic analysis, one can show that
\begin{equation}
 \frac{a^*(b)t-bt+1-x_N(t)}{d^*(b)t^{1/3}}
\label{xNsc}
\end{equation}
with $N=\nu t + 2c^*(\nu) t^{2/3} \tau$ tends to the Airy process where 
\begin{align}
 a^*(\nu) &=  (1-q)-(1-2q)\nu-2\sqrt{\nu(1-\nu)q(1-q)}, \\
 d^*(\nu) &= (\nu(\nu-1))^{\frac16}q^{\frac12}(1-q)^{\frac12}
             \left(1+\sqrt{\frac{\nu(1-q)}{q(\nu-1)}}\right)^{\frac23}
             \left(\sqrt{\frac{\nu-1}{\nu}}
                  -\sqrt{\frac{q}{1-q}}\right)^{\frac23}, \\
 c^*(\nu) &= (\nu(\nu-1))^{\frac56}
             \left(1+\sqrt{\frac{\nu(1-q)}{q(\nu-1)}}\right)^{\frac13}
             \left(\sqrt{\frac{\nu-1}{\nu}}
             -\sqrt{\frac{q}{1-q}}\right)^{\frac13}. 
\label{a_d}
\end{align}

In the cTASEP limit, (\ref{xNsc}) with (\ref{a_d}) replaced by 
\begin{align}
 a^*(\nu) &= (\sqrt{\nu}-1)^2, \\
 d^*(\nu) &= \nu^{1/6}(\sqrt{1/\nu}-1)^{2/3}, \\
 c^*(\nu) &= \nu^{\frac56} (\sqrt{1/\nu}-1)^{1/3} 
\end{align}
tends to the Airy process.

What have been explained in this subsection is the equal-time 
multi-particle joint distribution. 
It is also possible to study the multi-time joint distribution 
of the position of a single particle (tagged particle)
\cite{IS2007p}.

\setcounter{equation}{0}
\section{Green's function method}
So far we have explained how the fluctuation is 
computed by combining a combinatorial argument 
and techniques from random matrix theory. There is a related but 
a somehow different method using the Green's 
function for the TASEP. This is the probability that 
the particles starting from sites $y_1,\ldots,y_N$ 
($y_1>y_2>\ldots>y_N$) at time 0 are on sites 
$x_1,\ldots,x_N$ ($x_1>x_2>\ldots>x_N$) at time $t$
and is also called the transition probability.
We denote this quantity by 
$G(x_1,\ldots,x_N;t|y_1,\ldots,y_N;0)$ or, 
when we fix the initial condition, by $G(x_1,\ldots,x_N;t)$. 
For the cTASEP, a determinantal formula 
was obtained by Sch\"utz \cite{Schuetz1997b}. 
In this section we explain some applications 
of the formula. 
In this section we only consider the continuous time case. 
For the discrete case the expression for the Green's 
function was obtained in \cite{PP2006p} but has not been 
exploited to obtain the results in the scaling limit. 
See also \cite{BFP2006p}.

\subsection{Formula}
First let us remember the formula \cite{Schuetz1997b}.
It reads
\begin{equation}
G(x_1,x_2,\cdots,x_N;t) 
=
\det[F_{l-j}(x_{N-l+1}-y_{N-j+1};t)]_{j,l=1,\cdots,N} 
\label{G}
\end{equation}
where the function $F_n(x,t)$ appearing as matrix elements 
of the determinant is explicitly given by
\begin{equation}
 F_n(x;t)
 =
 e^{-t}\frac{t^x}{x!}\sum_{k=0}^{\infty} (-1)^k \frac{(n)_k}{(x+1)_k}
 \frac{t^k}{k!} .
\end{equation}
To see the structure we write down the determinants for $N=1,2$.
For $N=1$ this is simply
\begin{equation}
 G(x_1;t|y_1;0) = F_0(x-y;t)  = \frac{t^{x-y}}{(x-y)!}e^{-t}.
\end{equation}
For $N=2$ the formula gives 
\begin{equation}
 G(x_1,x_2;t) 
 =
 \begin{vmatrix}
  F_0(x_2-y_2;t)    & F_1(x_1-y_2;t) \\
  F_{-1}(x_2-y_1;t) & F_0(x_1-y_1;t)
 \end{vmatrix}.
\end{equation}
The proof of (\ref{G}) consists of checking the master 
equation and the initial condition 
which should be satisfied by the Green's function.

Some formulas of the function $F_n(x;t)$ which are useful 
in the following are given.  
It has an integral representation,
\begin{equation}
 F_{-n}(x;t) 
 =
 \frac{1}{2\pi i}\int_{\Gamma_{0,-1}} dw \frac{w^n}{(1+w)^{n+x+1}}e^{wt}
\end{equation}
where $\Gamma_{0,-1}$ represents a contour enclosing both 
$w=0$ and $w=-1$ anticlockwise. 
They satisfy the recurrence relations,
\begin{gather}
 \frac{d}{dt}F_n(x;t) = F_n(x-1;t) -F_n(x;t), \\
 \int_0^t dt F_{n-1}(x-1;t) = F_n(x;t), \\
 F_n(x;t)    = \sum_{x_1=x}^{\infty} F_{n-1}(x_1;t), \\
 F_{n-1}(x;t)= F_n(x;t) -F_n(x+1;t).
\end{gather}

\subsection{Position fluctuation of a single particle}
Let us first study the fluctuation of the position of a particle. 
A basic observation is that the probability we are interested in 
is written as a sum of the Green's function,
\begin{equation}
 \mathbb{P}[x_N(t)-y_N \geq M]
 =
 \sum_{y_N+M\leq x_N<\cdots<x_1} 
 G(x_1,\cdots,x_N;t|y_1,\cdots,y_N;0).
\end{equation}
Here $\mathbb{P}$ indicates the measure of the TASEP for our 
current initial condition and 
the summation on RHS is over all $x_i$'s which satisfy the constraint 
$y_N+M\leq x_N<\cdots<x_1$. 
The next thing to do is to rewrite the RHS
to a form which is suitable for an asymptotic analysis. 
In \cite{NS2004,RS2005}, it was shown that the quantity can be 
written in the form of a multiple integral which reduces
to (\ref{lue}) for the step initial condition. 

In this subsection we state a similar result for the  
fluctuation of $x_N(t)$ for the step initial condition, 
i.e., for $y_j=1-j(j=1,\ldots,N)$. We have
\begin{align}
 \mathbb{P}[x_N(t)+N-1\geq M] 
 =
 \frac{1}{Z} 
 \sum_{x_1=M}^{\infty}\cdots \sum_{x_N=M}^{\infty}
 \prod_{1\leq j<l\leq N} (x_j-x_l)^2
 \prod_{j=1}^N \frac{t^{x_j} e^{-t}}{x_j!} 
\label{ChUE}
\end{align}
with 
\begin{equation}
 Z = t^{N(N-1)/2}\prod_{j=1}^N j!.
\label{ChZ}
\end{equation}
Since the orthogonal polynomials for the weight 
$\frac{t^xe^{-t}}{x!}$ is called the Charlier orthogonal 
polynomials\cite{KS1996p}, 
this expression could  be called the Charlier 
ensemble representation of the fluctuation of $x_N(t)$.
The proof is almost the same as in \cite{NS2004} and is 
omitted here. But to illustrate the method we give a
step-by-step calculation for the $N=2$ case. 
It proceeds as follows.
\begin{align}
 \mathbb{P}[x_2(t)\geq M-1]
 &=
 \sum_{M-1\leq x_2<x_1} G(x_1,x_2;t|0,-1;0) \notag\\
 &=
 \sum_{M-1\leq x_2<x_1} 
 \begin{vmatrix}
   F_0(x_2+1;t)  & F_1(x_1+1;t) \\
   F_{-1}(x_2;t) & F_0(x_1;t) 
 \end{vmatrix} \notag\\
 &=
 \sum_{M-1\leq x_2<x_1} \sum_{y_1=x_1}^{\infty} 
 \begin{vmatrix}
   F_0(x_2+1;t)  & F_0(y_1+1;t) \\
   F_{-1}(x_2;t) & F_{-1}(y_1;t) 
 \end{vmatrix}.
\end{align}
If we denote the determinant on the RHS by $f(x_2,y_1)$,
this is equal to
\begin{align}
 &\quad\sum_{x_2=M-1}^{\infty} \sum_{x_1=x_2+1}^{\infty}\sum_{y_1=x_1}^{\infty}
 f(x_2,y_1)
 =
 \sum_{x_2=M-1}^{\infty} \sum_{y_1=x_2+1}^{\infty}\sum_{x_1=x_2+1}^{y_1}
 f(x_2,y_1) \notag\\
 &=
 \sum_{x_2=M}^{\infty} \sum_{y_1=x_2+1}^{\infty} (y_1-x_2) f(x_2,y_1) 
 =
 \sum_{x_2=M}^{\infty} \sum_{x_1=x_2+1}^{\infty}(x_1-x_2) f(x_2,x_1).
\end{align}
Since
\begin{align}
 f(x_2,x_1) 
 &=
 \begin{vmatrix}
   F_0(x_2+1;t)            & F_0(x_1+1;t)  \\
   F_0(x_2;t)-F_0(x_2+1;t) & F_0(x_1;t)-F_0(x_1+1;t) 
 \end{vmatrix}  \notag\\
 &=
 \begin{vmatrix}
   \frac{t^{x_2+1}}{(x_2+1)!}e^{-t} & \frac{t^{x_1+1}}{(x_1+1)!}e^{-t}\\
   \frac{t^{x_2}}{x_1!}e^{-t} & \frac{t^{x_1}}{x_1!}e^{-t}\\    
 \end{vmatrix} 
 =
 \frac{t^{x_1+x_2+1} e^{-2t}}{(x_1+1)!(x_2+1)!} (x_1-x_2),
\end{align}
we have
\begin{align}
 \mathbb{P}[x_2(t) \geq M-1] 
 &=
 \sum_{x_2=M+1}^{\infty} \sum_{x_1=x_2+1}^{\infty} (x_2-x_1)^2
 \frac{t^{x_1+x_2+1}}{(x_1+1)!(x_2+1)!} e^{-2t}
 \notag\\
 &=
 \frac{1}{2 t}\sum_{x_1=M}^{\infty} \sum_{x_2=M}^{\infty} (x_2-x_1)^2
 \frac{t^{x_1+x_2}}{x_1!x_2!} e^{-2t}.
\end{align}
This is nothing but the expression (\ref{ChUE}), (\ref{ChZ})
for $N=2$.

\subsection{Position of particles: Step case}
In the previous subsection we considered the fluctuation 
of a single particle. 
To study the joint distribution, the original form of 
the Green's function is not very convenient. 
We rewrite it in a form such that the position of particles 
can be regarded as a dynamics of the first walker in a certain
vicious walk \cite{Sasamoto2005,BFPS2006p}.

Let us consider a determinantal weight,
\begin{equation}
\prod_{r=1}^{N-1} \det[\phi_{r,r+1}^G(x_j^r,x_l^{r+1})]_{j,l=1,\ldots,r+1}
\cdot \det[\psi_j^{(N)}(x_l^N)]_{\begin{subarray}{c}j=0,...,N-1\\
                                               l=1,...,N \end{subarray}},
\label{Gw}
\end{equation}
under the condition $x_1^2 < x_2^2,x_1^3<x_2^3<x_3^3,\cdots$ and 
the convention $\phi^G_{r,r+1}(x_{r+1}^r;x_j^{r+1})=1$ for 
$r=1,\ldots,N-1, j=1,\ldots,r+1$.
Here the functions $\psi_j^{(r)}(x)$ and $\phi^G_{r_1,r_2}(x,y)$ 
are defined by 
\begin{align}
 \psi_j^{(r)}(x)    
 = 
 (-1)^{r-1-j} F_{-r+1+j}(x-y_{j+1},t) 
 =
 \frac{(-1)^{r-1-j}}{2\pi i}\int_{\G_{0,-1}} dw
 \frac{w^{r-1-j}}{(1+w)^{x+r-j-y_{j+1}}}e^{wt}
\end{align}
and 
\begin{align}
 \phi^G_{r_1,r_2}(x_1,x_2)
 =
 \begin{bmatrix}
  x_1-x_2-1 \\ r_2-r_1-1
 \end{bmatrix}.
\end{align}
In particular 
\begin{equation}
 \phi^G_{r,r+1}(x_1,x_2) = \begin{cases}
		       1, & x_1>x_2, \\ 0, & x_1 \leq x_2.
		      \end{cases}
\end{equation}
This can be considered as a weight of a vicious walk(Fig. 10). 
A different feature from (\ref{w_dPNG}) and (\ref{w_dPNGd})
is that the number of particles is not fixed in (\ref{Gw}).

Let $\mathbb{P}_G$ denote the corresponding measure. Then we have
\begin{align}
 G(x_1,\cdots,x_N;t) 
 =
 \mathbb{P}_G[x_1^1=x_1,x_1^2=x_2,\ldots,x_1^N=x_N].
\end{align}
For $N=2$ this reads
\begin{align*}
 G(x_1,x_2;t)
 &=
 \begin{vmatrix}
  F_0(x_2-y_2;t)    & F_1(x_1-y_2;t) \\
  F_{-1}(x_2-y_1;t) & F_0(x_1-y_1;t) 
 \end{vmatrix} \\  
 &=
 \sum_{x_2^2(>x_1^2)}
 \begin{vmatrix}
  \phi^G_{1,2}(x_1^1,x_1^2) & \phi^G_{1,2}(x_1^1,x_2^2) \\
  1                 & 1
 \end{vmatrix}
 \begin{vmatrix}
  \psi_0^{(2)}(x_1^2) & \psi_0^{(2)}(x_2^2) \\
  \psi_1^{(2)}(x_1^2) & \psi_1^{(2)}(x_2^2) 
 \end{vmatrix} 
\end{align*}
where $x_1^1=x_1,x_1^2=x_2$ and can be proved easily
by using properties the determinant and $F_n(x;t)$.
The proof for general $N$ is similar
and is given in \cite{BFPS2006p}.
The merit of rewriting this way is that the weight (\ref{Gw}) 
has a similar form to (\ref{GUE2t}) for tGUE. 
As a consequence we can describe the 
joint distribution by a Fredholm determinant. 
The kernel takes the form, 
\begin{equation}
 K_{N2}^G(r_1,x_1;r_2,x_2)
 =
 \tilde{K}_{N2}^G(r_1,x_1;r_2,x_2)
 -
 \phi_{r_1,r_2}^G(x_1,x_2)
\end{equation}
with 
\begin{equation}
\label{tK}
 \tilde{K}_{N2}^G(r_1,x_1;r_2,x_2)
 =
 \sum_{j=0}^{N-1} \psi_j^{(r_1)}(x_1) \varphi_j^{(r_2)}(x_2) 
\end{equation}
where $\varphi_j^{(r)}$ is a polynomial orthonormal to 
$\psi_j^{(r)}$.
For the step initial condition for which $y_{j+1}=-j$,
$\varphi_{j}^{(r)}$ ($j=0,1,\ldots,N-1$) can be taken as
\begin{equation}
 \vp_j^{(r)}(x)
 =
 \frac{(-1)^{r-1-j}}{2\pi i}\int_{\G_0}dz 
 \frac{(1+z)^{x+r-1}}{z^{r-j}}e^{-zt},
\end{equation}
where $\G_0$ is the contour enclosing the origin anticlockwise.
As a result we have a double integral expression of the kernel,
\begin{align}
 K_{N2}^G(r_1,x_1;r_2,x_2)
 &=
 \tilde{K}_{N2}^G(r_1,x_1;r_2,x_2) - \phi^G_{r_1,r_2}(x_1,x_2), \notag\\
 \tilde{K}_{N2}^G(r_1,x_1;r_2,x_2)
 &=
 \int_{\G_{-1}}\frac{dw}{2\pi i} \int_{\G_0} \frac{dz}{2\pi i}
 \frac{(1+z)^{x_2+r_2-1} (-w)^{r_1} e^{(w-z)t} }
      {(1+w)^{x_1+r_1} (-z)^{r_2} (w-z)}.
\label{Kint2}
\end{align}
By doing the asymptotic analysis as usual, we can reproduce 
the result in subsection 6.5 for the cTASEP. 

\subsection{Position of particles: Alternating case}
In the previous two subsections we have seen that the 
method using the Green's function allows us to reproduce 
the results for the step initial condition, which were 
obtained by a combinatorial approach. 
A big advantage of the method of using the Green's function 
is that one can deal with arbitrary initial configuration of 
particles at least up to some stage of the analysis. 

In particular in \cite{Sasamoto2005,BFPS2006p} the joint distribution 
for the alternating case was analyzed.
For a technical reason we take the initial condition to be 
$y_{j+1}=-2j$ ($j=0,1,\ldots,N-1$). But if we take $t$ finite and $N$
large enough, deep inside the negative region $x(\leq 0)$ the dynamics
of particles is the same as that for the alternating initial condition.
Remember here that up to now the multi-layer PNG techniques have not 
been successfully applied to obtain the multi-point statistics
for this initial condition.
The discussions are the same as for the step case until
(\ref{tK}). The difference is that this time we employ a 
different set of $\varphi_j^{(r)}$'s which are orthonormal to 
$\psi_j^{(r)}$'s. 
The functions are found to be
\begin{equation}
 \vp_j^{(r)}(x)
 =
 \frac{(-1)^{r-1-j}}{2\pi i}\int_{\G_0}dz \frac{1+2z}{1+z}
 \frac{(1+z)^{x+r+j-1}}{z^{r-j}}e^{-zt}.
\end{equation}
Then the kernel becomes
\begin{align}
 K_{N1}^G(r_1,x_1;r_2,x_2)
 &=
 \tilde{K}_{N1}^G(r_1,x_1;r_2,x_2) - \phi^G_{r_1,r_2}(x_1,x_2), \notag\\
 \tilde{K}_{N1}^G(r_1,x_1;r_2,x_2)
 &=
 \int_{\G_{-1}}\frac{dw}{2\pi i} \int_{\G_0}\frac{dz}{2\pi i}
 \frac{(1+z)^{x_2+r_2-2}(-w)^{r_1}(1+2z)e^{(w-z)t}}
      {(1+w)^{x_1+r_1-1}(-z)^{r_2}(w-z)(1+w+z)} .
\label{Kint}
\end{align}
By applying the saddle point analysis, one gets a limiting 
kernel which is different from (\ref{eAiK}). 
Though (\ref{Kint}) looks similar to (\ref{Kint2}), 
the pole at $z=-1-w$ makes the situation completely different. 

\setcounter{equation}{0}
\section{Concluding Remarks}
In this paper we have discussed the applications of random 
matrix techniques to the study of fluctuations of the 
one-dimensional totally asymmetric simple exclusion process(TASEP).
Besides a brief review of the techniques, we have 
explained three approaches.
The first one is the combinatorial approach, the second 
is the discrete polynuclear growth model and the last 
one is the Green's function method. 
The point is that in all these approaches there appear 
some objects whose weight can be described by a product 
of determinants and that the same kind of weight 
appears in the random matrix theory.
We have mainly explained the methods and results for 
the step initial condition. Other cases including 
the alternating initial condition are also briefly 
explained.
We have also seen the relationship of the TASEP 
to the directed polymer in random medium and the 
two types of the polynuclear growth models. 

In spite of its many successful applications,
it has not yet been clear to what extent  
this method could be extended to study other 
properties of the ASEP.
In fact there are a lot of open questions. 
For instance the correlation of the current 
at the origin $N(t_1),N(t_2)$ at two times is 
a natural quantity to consider, but we don't 
know if it is handled by a generalization of 
the random matrix techniques at the moment. 
More studies with new ideas are certainly desired.

\section*{Acknowledgment}
The author would like to thank T. Tanemura, M. Katori and T. Imamura 
for fruitful discussions.
This work is partly supported by the Grant-in-Aid for Young 
Scientists (B), the Ministry of Education, Culture, Sports, Science and 
Technology, Japan.

\setcounter{equation}{0}
\renewcommand{\theequation}{\Alph{section}.\arabic{equation}}
\renewcommand{\thesection}{Appendix \Alph{section}}
\setcounter{section}{0}
\section{}
In this appendix we explain some definitions, notations and 
formulas related to the Young tableaux. 
The partition $\lambda=(\lambda_1,\lambda_2,\ldots)$ 
is a collection of nonnegative integers $\lambda_i$'s 
which satisfy $\lambda_1 \geq \lambda_2 \geq \ldots$
and is conveniently represented by a Young diagram. 
For instance the partition $\lambda=(4,3,1,1)$ 
corresponds the Young diagram $\yng(4,3,1,1)$. 
A semistandard Young tableau(SSYT) is a Young diagram 
with each box filled with a nonnegative integer with 
the condition that they are non-decreasing in each 
row and are increasing in each column. We often restrict
the entries of integers to fill to be from a finite 
set $\{1,2,\ldots,N\}$.

The Schur function $s_{\lambda}(x_1,x_2,\ldots)$ is 
defined by 
\begin{equation}
 s_{\lambda}(x_1,x_2,\ldots) 
 = 
 \sum_{T:\text{SSYT}} x_1^{T_1} x_2^{T_2} \ldots
\end{equation}
where $T_i$ is the number of $i$ in SSYT. 
For instance when $\lambda=(2,1)$, $x_4=x_5=\ldots=0$
and we take entries from $\{1,2,3\}$, 
all possible SSYTs are 
\begin{equation}
 \young(11,2),~~ \young(11,3),~~ \young(12,2),~~ \young(12,3),~~ 
 \young(13,2),~~ \young(13,3),~~ \young(22,3),~~ \young(23,3)
\end{equation}
and hence
\begin{equation}
 s_{\lambda}(x_1,x_2,x_3,0,\ldots) 
 =
 x_1^2x_2 + x_1^2 x_3 + x_1 x_2^2 + x_1 x_3^2 
 + 
 x_2^2 x_3 + x_2 x_3^2 + 2 x_1 x_2 x_3. 
\end{equation}
The number of the SSYT's of shape $\lambda$ with entries
from $\{1,2,\ldots,N\}$ is given by 
\begin{equation}
 L(\lambda,N) = s_{\lambda}(\underbrace{1,\ldots,1}_N,0,\ldots).
\label{Ls}
\end{equation}
There is another representation of the Schur function,
\begin{equation}
 s_{\lambda}(x_1,\ldots,x_N,0,\ldots) 
 =
 \frac{\det(x_j^{\lambda_i+N-i})_{1\leq i,j \leq N}}
      {\det(x_j^{N-i})_{1 \leq i,j \leq N}} .
\end{equation}
By putting $x_i=\alpha^{i-1}$ ($i=1,\ldots,N$) and 
then taking the limit $\alpha\to 1$, one gets a formula, 
\begin{equation}
 s_{\lambda}(\underbrace{1,\ldots,1}_N,0,\ldots) 
 = 
 \prod_{1\leq i<j\leq N}\frac{\lambda_i-\lambda_j+j-i}{j-i}.
\label{sp}
\end{equation}


\newpage
\begin{large}
\noindent
Figure Captions
\end{large}

\vspace{10mm}
\noindent
Fig. 1:
The one-dimensional asymmetric simple exclusion 
process with parallel update. 

\vspace{10mm}
\noindent
Fig. 2:
The step initial condition for the TASEP. 

\vspace{10mm}
\noindent
Fig. 3:
An example of space-time trajectories of particles. 
The trajectories of the rightmost four particles are 
shown until each particle has made the fourth hops. 

\vspace{10mm}
\noindent
Fig. 4:
The dynamics of the eigenvalues of the GUE Dyson's Brownian 
motion (tGUE).

\vspace{10mm}
\noindent
Fig. 5:
The waiting time table. The time coordinate $s$ and 
space coordinate $x$ in the discrete PNG model are also 
shown. 

\vspace{10mm}
\noindent
Fig. 6:
The discrete PNG model. The second layer in the multi-layer 
PNG model is also shown. Initially at time $s=0$ the surface is 
flat and nothing happens until $s=1$. At time $s=1$ a nucleation
of height one occurs at the origin. Between time $s=1$ and $s=2$
the surface grows laterally in both directions with unit speed. 
The height at the origin at $s=3^-$ is determined by the 
``higher one wins'' rule. The height difference 1 becomes 
the height of a nucleation at the second layer. 

\vspace{10mm}
\noindent
Fig. 7:
A snapshot of a simulation of the multi-layer discrete PNG model.
Note a similarity to the Fig. 4. 

\vspace{10mm}
\noindent
Fig. 8:
The alternating initial condition for the TASEP. 

\vspace{10mm}
\noindent
Fig. 9:
Another discrete PNG model in subsection 6.5. 
The height of the top layer gives the positions of 
particles through (\ref{xk}). Nucleations at a 
lower layer occur stochastically. 

\vspace{10mm}
\noindent
Fig. 10:
Vicious walk related to the Green's function of TASEP. 
The height of the top layer gives the positions of particles.
The dependence on the initial condition is reflected in 
the function $\psi_j(x)$.


\newpage
\renewcommand{\thepage}{Figure 1}

\unitlength .6mm
\begin{picture}(160,40)(-8,5)
\multiput(0,10)(0.5,0){20}{\line(1,0){0.3}}
\multiput(0,30)(0.5,0){20}{\line(1,0){0.3}}
\multiput(150,10)(0.5,0){20}{\line(1,0){0.3}}
\multiput(150,30)(0.5,0){20}{\line(1,0){0.3}}
\put(10,10){\line(1,0){140}}
\put(10,30){\line(1,0){140}}
\put(10,10){\line(0,1){20}}
\put(30,10){\line(0,1){20}}
\put(50,10){\line(0,1){20}}
\put(70,10){\line(0,1){20}}
\put(90,10){\line(0,1){20}}
\put(110,10){\line(0,1){20}}
\put(130,10){\line(0,1){20}}
\put(150,10){\line(0,1){20}}
\put(0,19){$\cdots$}
\put(40,20){\circle*{10}}
\put(46,18){{\Large $\Rightarrow$}}
\put(45,33){$1-q$}
\put(80,20){\circle*{10}}
\put(100,20){\circle*{10}}
\put(106,18){{\Large $\Rightarrow$}}
\put(105,33){$1-q$}
\put(140,20){\circle*{10}}
\put(153,19){$\cdots$}
\put(17,0){-3}
\put(37,0){-2}
\put(57,0){-1}
\put(79,0){0}
\put(99,0){1}
\put(119,0){2}
\put(139,0){3}

\end{picture}

\newpage
\renewcommand{\thepage}{Figure 2}

\unitlength 0.6mm
\begin{picture}(160,35)(-8,0)
\multiput(0,10)(0.5,0){20}{\line(1,0){0.3}}
\multiput(0,30)(0.5,0){20}{\line(1,0){0.3}}
\multiput(150,10)(0.5,0){20}{\line(1,0){0.3}}
\multiput(150,30)(0.5,0){20}{\line(1,0){0.3}}
\put(10,10){\line(1,0){140}}
\put(10,30){\line(1,0){140}}
\put(10,10){\line(0,1){20}}
\put(30,10){\line(0,1){20}}
\put(50,10){\line(0,1){20}}
\put(70,10){\line(0,1){20}}
\put(90,10){\line(0,1){20}}
\put(110,10){\line(0,1){20}}
\put(130,10){\line(0,1){20}}
\put(150,10){\line(0,1){20}}
\put(0,19){$\cdots$}
\put(20,20){\circle*{10}}
\put(40,20){\circle*{10}}
\put(60,20){\circle*{10}}
\put(80,20){\circle*{10}}
\put(17,0){-3}
\put(37,0){-2}
\put(57,0){-1}
\put(79,0){0}
\put(99,0){1}
\put(119,0){2}
\put(139,0){3}
\end{picture}

\newpage
\renewcommand{\thepage}{Figure 3}

\begin{figure}[h]
\unitlength .8mm
\begin{picture}(400,200)
\put(10,20){\line(1,0){80}}
\multiput(10,30)(1,0){80}{\line(1,0){0.2}}
\multiput(10,40)(1,0){80}{\line(1,0){0.2}}
\multiput(10,50)(1,0){80}{\line(1,0){0.2}}
\multiput(10,60)(1,0){80}{\line(1,0){0.2}}
\multiput(10,70)(1,0){80}{\line(1,0){0.2}}
\multiput(10,80)(1,0){80}{\line(1,0){0.2}}
\multiput(10,90)(1,0){80}{\line(1,0){0.2}}
\multiput(10,100)(1,0){80}{\line(1,0){0.2}}
\multiput(10,110)(1,0){80}{\line(1,0){0.2}}
\multiput(10,120)(1,0){80}{\line(1,0){0.2}}
\multiput(10,130)(1,0){80}{\line(1,0){0.2}}
\multiput(10,140)(1,0){80}{\line(1,0){0.2}}
\multiput(10,150)(1,0){80}{\line(1,0){0.2}}
\multiput(10,160)(1,0){80}{\line(1,0){0.2}}
\multiput(10,170)(1,0){80}{\line(1,0){0.2}}

\put(10,20){\line(0,1){150}}
\multiput(15,20)(0,1){150}{\line(1,0){0.2}}
\multiput(25,20)(0,1){150}{\line(1,0){0.2}}
\multiput(35,20)(0,1){150}{\line(1,0){0.2}}
\multiput(45,20)(0,1){150}{\line(1,0){0.2}}
\multiput(55,20)(0,1){150}{\line(1,0){0.2}}
\multiput(65,20)(0,1){150}{\line(1,0){0.2}}
\multiput(75,20)(0,1){150}{\line(1,0){0.2}}
\multiput(85,20)(0,1){150}{\line(1,0){0.2}}

\put(10,20){\line(0,1){40}}
\put(12,10){-3}
\put(22,10){-2}
\put(32,10){-1}
\put(44,10){0}
\put(54,10){1}
\put(64,10){2}
\put(74,10){3}
\put(84,10){4}

\put(4,18){0}
\put(4,28){1}
\put(4,38){2}
\put(4,48){3}
\put(4,58){4}
\put(4,68){5}
\put(4,78){6}
\put(4,88){7}
\put(4,98){8}
\put(4,108){9}
\put(3,118){10}
\put(3,128){11}
\put(3,138){12}
\put(3,148){13}
\put(3,158){14}
\put(3,168){15}
\put(0,170){$t$}
\put(90,15){$x$}

\put(45,20){\line(0,1){10}}
\put(45,30){\line(1,1){10}}
\put(55,40){\line(0,1){20}}
\put(55,60){\line(1,1){10}}
\put(65,70){\line(0,1){10}}
\put(65,80){\line(1,1){10}}
\put(75,90){\line(0,1){10}}
\put(75,100){\line(1,1){10}}
\put(45,20){\circle*{3}}
\put(45,30){\circle*{3}}
\put(55,40){\circle*{3}}
\put(55,50){\circle*{3}}
\put(55,60){\circle*{3}}
\put(65,70){\circle*{3}}
\put(65,80){\circle*{3}}
\put(75,90){\circle*{3}}
\put(75,100){\circle*{3}}
\put(85,110){\circle*{3}}

\put(35,20){\line(0,1){30}}
\put(35,50){\line(1,1){10}}
\put(45,60){\line(0,1){30}}
\put(45,90){\line(1,1){20}}
\put(65,110){\line(0,1){20}}
\put(65,130){\line(1,1){10}}
\put(35,20){\circle*{3}}
\put(35,30){\circle*{3}}
\put(35,40){\circle*{3}}
\put(35,50){\circle*{3}}
\put(45,60){\circle*{3}}
\put(45,70){\circle*{3}}
\put(45,80){\circle*{3}}
\put(45,90){\circle*{3}}
\put(55,100){\circle*{3}}
\put(65,110){\circle*{3}}
\put(65,120){\circle*{3}}
\put(65,130){\circle*{3}}
\put(75,140){\circle*{3}}

\put(25,20){\line(0,1){50}}
\put(25,70){\line(1,1){10}}
\put(35,80){\line(0,1){30}}
\put(35,110){\line(1,1){20}}
\put(55,130){\line(0,1){20}}
\put(55,150){\line(1,1){10}}
\put(25,20){\circle*{3}}
\put(25,30){\circle*{3}}
\put(25,40){\circle*{3}}
\put(25,50){\circle*{3}}
\put(25,60){\circle*{3}}
\put(25,70){\circle*{3}}
\put(35,80){\circle*{3}}
\put(35,90){\circle*{3}}
\put(35,100){\circle*{3}}
\put(35,110){\circle*{3}}
\put(45,120){\circle*{3}}
\put(55,130){\circle*{3}}
\put(55,140){\circle*{3}}
\put(55,150){\circle*{3}}
\put(65,160){\circle*{3}}

\put(15,20){\line(0,1){90}}
\put(15,110){\line(1,1){20}}
\put(35,130){\line(0,1){10}}
\put(35,140){\line(1,1){10}}
\put(45,150){\line(0,1){10}}
\put(45,160){\line(1,1){10}}
\put(15,20){\circle*{3}}
\put(15,30){\circle*{3}}
\put(15,40){\circle*{3}}
\put(15,50){\circle*{3}}
\put(15,60){\circle*{3}}
\put(15,70){\circle*{3}}
\put(15,80){\circle*{3}}
\put(15,90){\circle*{3}}
\put(15,100){\circle*{3}}
\put(15,110){\circle*{3}}
\put(25,120){\circle*{3}}
\put(35,130){\circle*{3}}
\put(35,140){\circle*{3}}
\put(45,150){\circle*{3}}
\put(45,160){\circle*{3}}
\put(55,170){\circle*{3}}

\end{picture}
\end{figure}

\newpage
\renewcommand{\thepage}{Figure 4}

\begin{picture}(400,300)
\psfrag{t}{$t$}
\psfrag{x}{$x$}
\psfrag{t_1}{$t_1$}
\psfrag{t_2}{$t_2$}
\put(50,100){\includegraphics{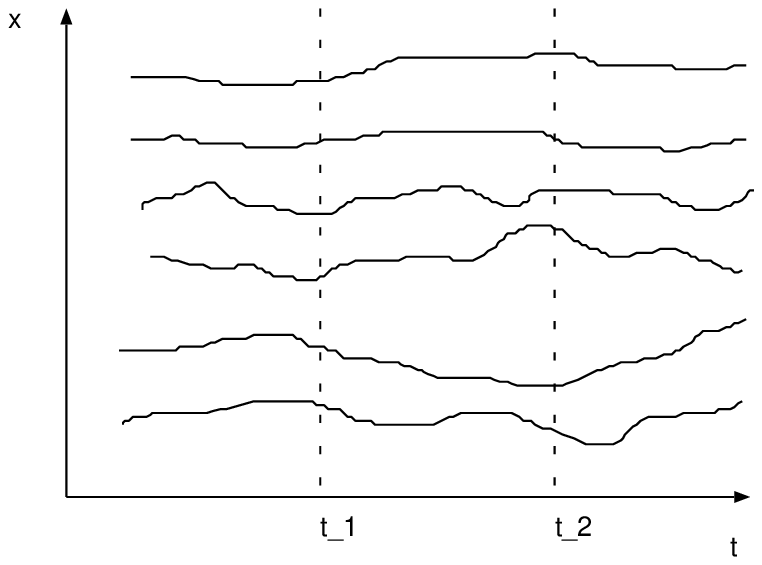}}
\end{picture}

\newpage
\renewcommand{\thepage}{Figure 5}

 \unitlength 0.8mm
  \begin{picture}(160,50)(-20,25)
   \put(0,70){\vector(1,0){50}}
   \put(0,70){\vector(0,-1){50}}
   \put(0,60){\line(1,0){50}}
   \put(0,50){\line(1,0){50}}
   \put(0,40){\line(1,0){50}}
   \put(0,30){\line(1,0){50}}
   \put(10,70){\line(0,-1){50}}
   \put(20,70){\line(0,-1){50}}
   \put(30,70){\line(0,-1){50}}
   \put(40,70){\line(0,-1){50}}
   \put(4,64){$1$}
   \put(14,64){$1$}
   \put(24,64){$1$}
   \put(34,64){$3$}
   \put(44,64){$\cdots$}
   \put(4,54){$2$}
   \put(14,54){$2$}
   \put(24,54){$1$}
   \put(34,54){$0$}
   \put(4,44){$1$}
   \put(14,44){$0$}
   \put(24,44){$0$}
   \put(34,44){$1$}
   \put(4,34){$1$}
   \put(14,34){$2$}
   \put(24,34){$4$}
   \put(34,34){$0$}
   \put(4,24){$\vdots$}

   \put(-5,15){$i$}
   \put(55,70){$j$}
   \put(80,70){\vector(-1,-1){10}}
   \put(80,70){\vector(1,-1){10}}
   \put(68,55){$x$}
   \put(88,55){$s$}
  \end{picture}

\newpage
\renewcommand{\thepage}{Figure 6}

\begin{figure}[h]
\unitlength .8mm
\begin{picture}(400,60)
\put(10,20){\line(1,0){60}}
\put(15,20){\circle*{3}}
\put(25,20){\circle*{3}}
\put(35,20){\circle*{3}}
\put(45,20){\circle*{3}}
\put(55,20){\circle*{3}}
\put(65,20){\circle*{3}}

\put(10,30){\line(1,0){60}}
\put(15,30){\circle*{3}}
\put(25,30){\circle*{3}}
\put(35,30){\circle*{3}}
\put(45,30){\circle*{3}}
\put(55,30){\circle*{3}}
\put(65,30){\circle*{3}}

\put(10,20){\line(0,1){40}}
\put(12,10){-2}
\put(22,10){-1}
\put(34,10){0}
\put(44,10){1}
\put(54,10){2}
\put(64,10){3}
\put(5,18){-1}
\put(5,28){0}
\put(5,38){1}
\put(5,48){2}
\put(5,58){$h$}
\put(34,62){$s=0$}

\put(80,20){\line(1,0){60}}
\put(85,20){\circle*{3}}
\put(95,20){\circle*{3}}
\put(105,20){\circle*{3}}
\put(115,20){\circle*{3}}
\put(125,20){\circle*{3}}
\put(135,20){\circle*{3}}

\put(80,30){\line(1,0){60}}
\put(85,30){\circle*{3}}
\put(95,30){\circle*{3}}
\put(105,30){\circle*{3}}
\put(115,30){\circle*{3}}
\put(125,30){\circle*{3}}
\put(135,30){\circle*{3}}

\put(82,10){-2}
\put(92,10){-1}
\put(104,10){0}
\put(114,10){1}
\put(124,10){2}
\put(134,10){3}
\put(102,62){$s=1^{-}$}

\put(143,15){$x$}

\end{picture}
\end{figure}

\begin{figure}[h]
\unitlength .8mm
\begin{picture}(400,60)
\put(10,30){\line(1,0){25}}
\put(35,30){\line(0,1){10}}
\put(35,40){\line(1,0){10}}
\put(45,40){\line(0,-1){10}}
\put(45,30){\line(1,0){25}}
\put(15,30){\circle*{3}}
\put(25,30){\circle*{3}}
\put(35,40){\circle*{3}}
\put(45,30){\circle*{3}}
\put(55,30){\circle*{3}}
\put(65,30){\circle*{3}}

\put(15,20){\circle*{3}}
\put(10,20){\line(1,0){60}}
\put(25,20){\circle*{3}}
\put(35,20){\circle*{3}}
\put(45,20){\circle*{3}}
\put(55,20){\circle*{3}}
\put(65,20){\circle*{3}}

\put(10,20){\line(0,1){40}}
\put(12,10){-2}
\put(22,10){-1}
\put(34,10){0}
\put(44,10){1}
\put(54,10){2}
\put(64,10){3}
\put(5,18){-1}
\put(5,28){0}
\put(5,38){1}
\put(5,48){2}
\put(5,58){$h$}
\put(34,62){$s=1$}

\put(80,20){\line(1,0){60}}
\put(85,20){\circle*{3}}
\put(95,20){\circle*{3}}
\put(105,20){\circle*{3}}
\put(115,20){\circle*{3}}
\put(125,20){\circle*{3}}
\put(135,20){\circle*{3}}

\put(80,30){\line(1,0){15}}
\put(95,30){\line(0,1){10}}
\put(95,40){\line(1,0){30}}
\put(125,40){\line(0,-1){10}}
\put(125,30){\line(1,0){15}}
\put(85,30){\circle*{3}}
\put(95,40){\circle*{3}}
\put(105,40){\circle*{3}}
\put(115,40){\circle*{3}}
\put(125,30){\circle*{3}}
\put(135,30){\circle*{3}}

\put(82,10){-2}
\put(92,10){-1}
\put(104,10){0}
\put(114,10){1}
\put(124,10){2}
\put(134,10){3}
\put(102,62){$s=2^{-}$}

\put(143,15){$x$}
\end{picture}
\end{figure}

\begin{figure}[h]
\unitlength .8mm
\begin{picture}(400,60)
\put(10,10){\line(1,0){60}}
\put(15,10){\circle*{3}}
\put(25,10){\circle*{3}}
\put(35,10){\circle*{3}}
\put(45,10){\circle*{3}}
\put(55,10){\circle*{3}}
\put(65,10){\circle*{3}}

\put(10,20){\line(1,0){15}}
\put(25,20){\line(0,1){20}}
\put(25,40){\line(1,0){10}}
\put(35,40){\line(0,-1){10}}
\put(35,30){\line(1,0){10}}
\put(45,30){\line(0,1){20}}
\put(45,50){\line(1,0){10}}
\put(55,50){\line(0,-1){30}}
\put(55,20){\line(1,0){15}}
\put(15,20){\circle*{3}}
\put(25,40){\circle*{3}}
\put(35,30){\circle*{3}}
\put(45,50){\circle*{3}}
\put(55,20){\circle*{3}}
\put(65,20){\circle*{3}}

\put(10,10){\line(0,1){50}}
\put(12,0){-2}
\put(22,0){-1}
\put(34,0){0}
\put(44,0){1}
\put(54,0){2}
\put(64,0){3}
\put(3,8){-1}
\put(5,18){0}
\put(5,28){1}
\put(5,38){2}
\put(5,48){3}
\put(5,58){$h$}
\put(34,62){$s=2$}

\put(80,10){\line(1,0){25}}
\put(105,10){\line(0,1){10}}
\put(105,20){\line(1,0){10}}
\put(115,20){\line(0,-1){10}}
\put(115,10){\line(1,0){25}}
\put(85,10){\circle*{3}}
\put(95,10){\circle*{3}}
\put(105,20){\circle*{3}}
\put(115,10){\circle*{3}}
\put(125,10){\circle*{3}}
\put(135,10){\circle*{3}}

\put(80,20){\line(1,0){5}}
\put(85,20){\line(0,1){20}}
\put(85,40){\line(1,0){20}}
\put(105,40){\line(0,1){10}}
\put(105,50){\line(1,0){30}}
\put(135,50){\line(0,-1){30}}
\put(135,20){\line(1,0){5}}
\put(85,40){\circle*{3}}
\put(95,40){\circle*{3}}
\put(105,50){\circle*{3}}
\put(115,50){\circle*{3}}
\put(125,50){\circle*{3}}
\put(135,20){\circle*{3}}

\put(82,0){-2}
\put(92,0){-1}
\put(104,0){0}
\put(114,0){1}
\put(124,0){2}
\put(134,0){3}
\put(102,62){$s=3^{-}$}

\put(143,5){$x$}
\end{picture}
\end{figure}

\newpage
\renewcommand{\thepage}{Figure 7}

\begin{picture}(400,150)
\put(150,5){$x$}
\put(-5,110){$h$}
\put(0,10){\includegraphics{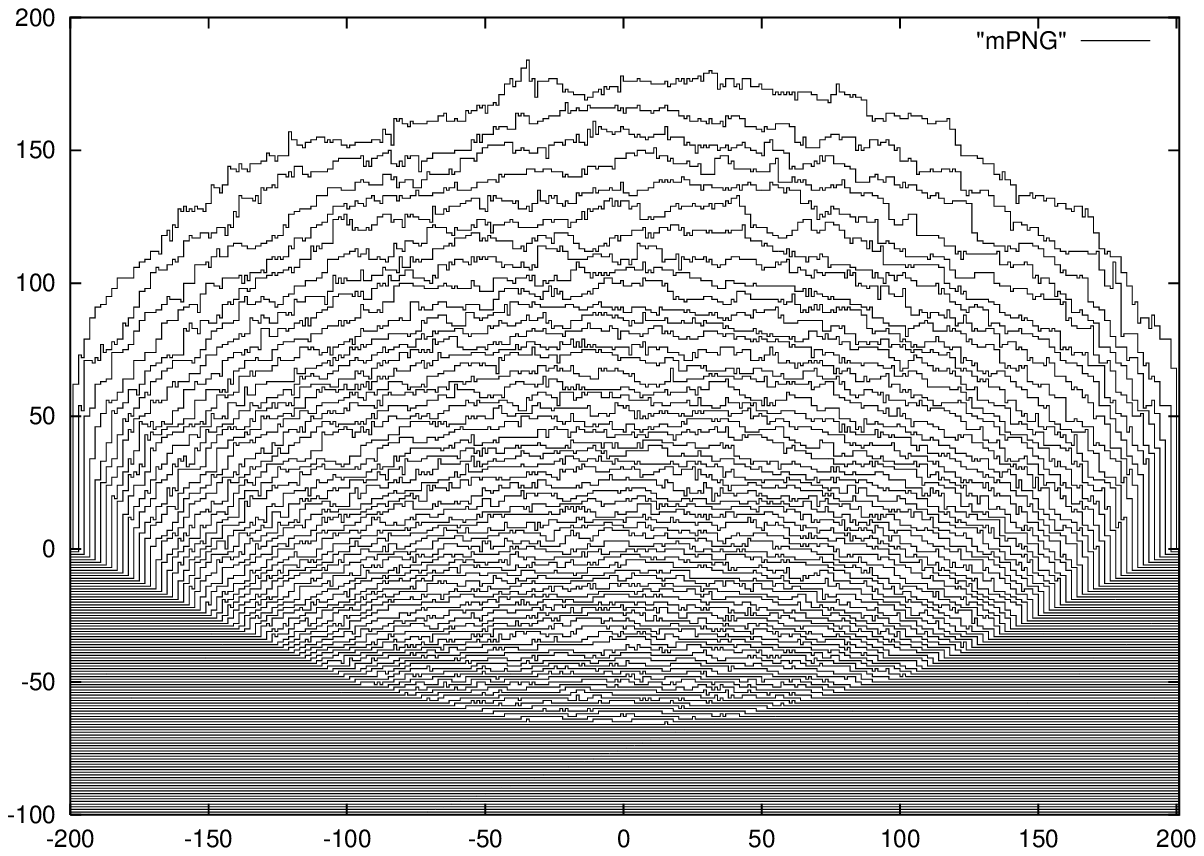}}
\end{picture}

\newpage
\renewcommand{\thepage}{Figure 8}

\unitlength 0.6mm
\begin{picture}(160,35)(-8,0)
\multiput(0,10)(0.5,0){20}{\line(1,0){0.3}}
\multiput(0,30)(0.5,0){20}{\line(1,0){0.3}}
\multiput(150,10)(0.5,0){20}{\line(1,0){0.3}}
\multiput(150,30)(0.5,0){20}{\line(1,0){0.3}}
\put(10,10){\line(1,0){140}}
\put(10,30){\line(1,0){140}}
\put(10,10){\line(0,1){20}}
\put(30,10){\line(0,1){20}}
\put(50,10){\line(0,1){20}}
\put(70,10){\line(0,1){20}}
\put(90,10){\line(0,1){20}}
\put(110,10){\line(0,1){20}}
\put(130,10){\line(0,1){20}}
\put(150,10){\line(0,1){20}}
\put(0,19){$\cdots$}
\put(40,20){\circle*{10}}
\put(80,20){\circle*{10}}
\put(120,20){\circle*{10}}
\put(153,19){$\cdots$}
\put(17,0){-3}
\put(37,0){-2}
\put(57,0){-1}
\put(79,0){0}
\put(99,0){1}
\put(119,0){2}
\put(139,0){3}

\end{picture}

\newpage
\renewcommand{\thepage}{Figure 9}

\begin{figure}[h]
\unitlength .8mm
\begin{picture}(400,55)
\put(34,60){$t=0$}

\put(10,20){\line(1,0){60}}
\put(10,20){\circle*{3}}
\put(20,20){\circle*{3}}
\put(30,20){\circle*{3}}
\put(40,20){\circle*{3}}
\put(50,20){\circle*{3}}
\put(60,20){\circle*{3}}
\put(70,20){\circle*{3}}

\put(10,30){\line(1,0){60}}
\put(10,30){\circle*{3}}
\put(20,30){\circle*{3}}
\put(30,30){\circle*{3}}
\put(40,30){\circle*{3}}
\put(50,30){\circle*{3}}
\put(60,30){\circle*{3}}
\put(70,30){\circle*{3}}

\put(4,58){$k$}
\put(10,20){\line(0,1){40}}
\put(2,18){-1}
\put(4,28){0}
\put(4,38){1}
\put(4,48){2}
\end{picture}
\end{figure}

\begin{figure}[h]
\unitlength .8mm
\begin{picture}(400,45)
\put(34,60){$t=1$}

\put(10,20){\line(1,0){60}}
\put(10,20){\circle*{3}}
\put(20,20){\circle*{3}}
\put(30,20){\circle*{3}}
\put(40,20){\circle*{3}}
\put(50,20){\circle*{3}}
\put(60,20){\circle*{3}}
\put(70,20){\circle*{3}}

\put(10,30){\line(0,1){10}}
\put(10,40){\line(1,0){10}}
\put(20,40){\line(1,-1){10}}
\put(30,30){\line(1,0){40}}

\put(10,30){\circle*{3}}
\put(20,40){\circle*{3}}
\put(30,30){\circle*{3}}
\put(40,30){\circle*{3}}
\put(50,30){\circle*{3}}
\put(60,30){\circle*{3}}
\put(70,30){\circle*{3}}

\put(4,58){$k$}
\put(10,20){\line(0,1){40}}
\put(2,18){-1}
\put(4,28){0}
\put(4,38){1}
\put(4,48){2}
\end{picture}
\end{figure}

\begin{figure}[h]
\unitlength .8mm
\begin{picture}(400,45)
\put(34,60){$t=2$}

\put(10,20){\line(1,0){20}}
\put(30,20){\line(0,1){10}}
\put(30,30){\line(1,0){10}}
\put(40,30){\line(1,-1){10}}
\put(50,20){\line(1,0){20}}
\put(10,20){\circle*{3}}
\put(20,20){\circle*{3}}
\put(30,20){\circle*{3}}
\put(40,30){\circle*{3}}
\put(50,20){\circle*{3}}
\put(60,20){\circle*{3}}
\put(70,20){\circle*{3}}

\put(10,30){\line(0,1){10}}
\put(10,40){\line(1,0){30}}
\put(40,40){\line(1,-1){10}}
\put(50,30){\line(1,0){20}}
\put(10,30){\circle*{3}}
\put(20,40){\circle*{3}}
\put(30,40){\circle*{3}}
\put(40,40){\circle*{3}}
\put(50,30){\circle*{3}}
\put(60,30){\circle*{3}}
\put(70,30){\circle*{3}}

\put(4,58){$k$}
\put(10,20){\line(0,1){40}}
\put(2,18){-1}
\put(4,28){0}
\put(4,38){1}
\put(4,48){2}
\end{picture}
\end{figure}

\begin{figure}[!h]
\unitlength .8mm
\begin{picture}(400,45)
\put(34,60){$t=3$}

\put(10,20){\line(1,0){20}}
\put(30,20){\line(0,1){10}}
\put(30,30){\line(1,0){30}}
\put(60,30){\line(1,-1){10}}
\put(10,20){\circle*{3}}
\put(20,20){\circle*{3}}
\put(30,20){\circle*{3}}
\put(40,30){\circle*{3}}
\put(50,30){\circle*{3}}
\put(60,30){\circle*{3}}
\put(70,20){\circle*{3}}

\put(10,30){\line(0,1){20}}
\put(10,50){\line(1,0){10}}
\put(20,50){\line(1,-1){10}}
\put(30,40){\line(0,1){10}}
\put(30,50){\line(1,0){10}}
\put(40,50){\line(1,-1){10}}
\put(50,40){\line(1,0){10}}
\put(60,40){\line(1,-1){10}}
\put(10,30){\circle*{3}}
\put(20,50){\circle*{3}}
\put(30,40){\circle*{3}}
\put(40,50){\circle*{3}}
\put(50,40){\circle*{3}}
\put(60,40){\circle*{3}}
\put(70,30){\circle*{3}}

\put(80,10){$v$}
\put(8,10){0}
\put(18,10){1}
\put(28,10){2}
\put(38,10){3}
\put(48,10){4}
\put(58,10){5}
\put(68,10){6}

\put(80,3){$r$}
\put(18,3){1}
\put(38,3){2}
\put(58,3){3}

\put(4,58){$k$}
\put(10,20){\line(0,1){40}}
\put(2,18){-1}
\put(4,28){0}
\put(4,38){1}
\put(4,48){2}

\end{picture}
\end{figure}

\newpage
\renewcommand{\thepage}{Figure 10}

\begin{picture}(200,120)(-10,)
\put(10,100){\vector(1,0){160}}
\put(10,98){\line(0,1){4}}
\put(8,105){$1$}
\put(25,110){$\phi^G$}
\put(40,98){\line(0,1){4}}
\put(38,105){$2$}
\put(55,110){$\phi^G$}
\put(70,98){\line(0,1){4}}
\put(68,105){$3$}
\put(95,105){$\cdots$}
\put(130,98){\line(0,1){4}}
\put(115,105){$N-1$}
\put(145,115){$\phi^G$}
\put(160,98){\line(0,1){4}}
\put(157,105){$N$}
\put(157,118){$\psi$}
\put(173,103){$r$}
\put(10,80){\circle*{5}}
\put(7,70){$x_1^1$}
\put(10,80){\line(6,1){30}}
\put(40,85){\circle*{5}}
\put(37,73){$x_1^2$}
\put(40,85){\line(10,-1){30}}
\put(70,82){\circle*{5}}
\put(67,70){$x_1^3$}
\put(70,82){\line(10,1){20}}
\put(95,71){$\cdots$}
\put(110,81){\line(20,-1){20}}
\put(130,80){\circle*{5}}
\put(125,68){$x_1^{N-1}$}
\put(130,80){\line(10,1){30}}
\put(160,83){\circle*{5}}
\put(155,73){$x_1^N$}
\put(40,60){\circle*{5}}
\put(35,48){$x_2^2$}
\put(40,60){\line(15,1){30}}
\put(70,62){\circle*{5}}
\put(65,50){$x_2^3$}
\put(70,62){\line(10,-1){20}}
\put(95,51){$\cdots$}
\put(110,60){\line(1,0){20}}
\put(130,60){\circle*{5}}
\put(125,47){$x_2^{N-1}$}
\put(130,60){\line(10,-1){30}}
\put(160,57){\circle*{5}}
\put(155,47){$x_2^N$}
\put(70,42){\circle*{5}}
\put(65,30){$x_3^3$}
\put(70,42){\line(10,1){20}}
\put(95,31){$\cdots$}
\put(110,40){\line(10,-1){20}}
\put(130,38){\circle*{5}}
\put(125,25){$x_3^{N-1}$}
\put(130,38){\line(30,1){30}}
\put(160,39){\circle*{5}}
\put(155,29){$x_3^N$}
\put(158,18){$\vdots$}
\put(160,10){\circle*{5}}
\put(155,0){$x_N^N$}
\end{picture}

\end{document}